\documentclass[12pt, preprint]{emulateapj}
\usepackage{amsmath}
\usepackage{graphicx}

\shortauthors{L. BOCO ET AL.}
\shorttitle{Evolution of galaxy star formation and metallicity: impact on double compact objects mergers}
\slugcomment{Accepted by ApJ}

\begin{document}

\title{Evolution of galaxy star formation and metallicity:\\ impact on double compact objects mergers}

\author{L. Boco\altaffilmark{1,2,3}, A. Lapi\altaffilmark{1,2,3,4}, M. Chruslinska\altaffilmark{5}, D. Donevski\altaffilmark{1}, A. Sicilia\altaffilmark{1}, L. Danese\altaffilmark{1,2}}
\altaffiltext{1}{SISSA, Via Bonomea 265, 34136 Trieste, Italy}\altaffiltext{2}{IFPU-Institute for Fundamental Physics of the Universe, Via Beirut 2, 34014 Trieste, Italy}\altaffiltext{3}{INFN-Sezione di Trieste, via Valerio 2, 34127 Trieste,  Italy}\altaffiltext{4}{INAF-Osservatorio Astronomico di Trieste, via Tiepolo 11, 34131 Trieste, Italy}\altaffiltext{5}{Department of Astrophysics/IMAPP, Radboud University, PO Box 9010, NL-6500 GL Nijmegen, The Netherlands}

\begin{abstract}
We study the impact of different galaxy statistics and empirical metallicity scaling relations on the merging rates and on the properties of compact objects binaries. First, we analyze the similarities and differences of using the star formation rate functions or the stellar mass functions as galaxy statistics for the computation of the cosmic star formation rate density. Then we investigate the effects of adopting the Fundamental Metallicity Relation or a classic Mass Metallicity Relation to assign metallicity to galaxies with given properties. We find that when the Fundamental Metallicity Relation is exploited, the bulk of the star formation occurs at relatively high metallicities even at high redshift; the opposite holds when the Mass Metallicity Relation is employed, since in this case the metallicity at which most of the star formation takes place strongly decreases with redshift. We discuss the various reasons and possible biases originating this discrepancy. Finally, we show the impact that these different astrophysical prescriptions have on the merging rates and on the properties of compact objects binaries; specifically, we present results for the redshift dependent merging rates and for the chirp mass and time delay distributions of the merging binaries.
\end{abstract}


\section{Introduction}
The discovery of gravitational waves (GWs) by the LIGO/Virgo team (Abbott et al. 2016a, 2016b, 2016c, 2017a, 2017b, 2017c, 2017d, 2017e, 2019; 2020a; 2020b; also https://www.ligo.org/) has opened an observational window on the Universe with a new messenger. On the one hand, even few GW events with detected electromagnetic counterparts can be of enormous importance for cosmology and fundamental physics (Creminelli et al. 2017; Radice et al. 2018; Yang et al. 2019); on the other hand, large statistics of GWs can yield many astrophysical information on stellar and binary evolution (e.g., Belczynski et al. 2016; Dvorkin et al. 2018; Mapelli \& Giacobbo 2018), on the properties of the host galaxies such as chemical evolution, star formation histories, initial mass function (IMF; e.g., O’Shaughnessy et al. 2010; de Mink \& Belczynski 2015), and even on cosmology at large (e.g., Taylor \& Gair 2012; Nissanke et al. 2013; Liao et al. 2017; Fishbach et al. 2019). Moreover, the possibility to cross correlate the GW detected signals with some other tracers of the Large Scale Structures can help to improve cosmological constraints or to test competing astrophysical frameworks (e.g. Oguri 2016; Raccanelli et al. 2016; Scelfo et al. 2018, 2020; Calore et al. 2020). 

Given these numerous applications, it is important to well characterize the population of compact objects (COs) merging binaries, to compute the related merging rates and to understand their dependence on different astrophysical scenarios. The merging rates per unit volume and chirp mass $\rm \mathcal{M}$\footnote{The chirp mass is defined as: $\mathcal{M}\equiv (m_1\,m_2)^{3/5}/(m_1+m_2)^{1/5}$, where $m_1$ and $m_2$ are the masses of the two merging objects.} as a function of the cosmic time can be computed as (see Barrett et al. 2018 and Neijssel et al. 2019):
\begin{equation}
    \frac{\rm d\dot{N}}{\rm dVd\mathcal{M}}(\rm t)=\int\rm dt_{\rm d}\int\rm dZ\frac{\rm dN}{\rm dM_{\rm SFR}d\mathcal{M}dt_{\rm d}}(\rm Z)\frac{\rm d\dot{M}_{\rm SFR}}{\rm dVdZ}(t-t_{\rm d})
    \label{eq:merging rates}
\end{equation}
where $\rm t$ is the cosmic time, equivalent to redshift, $\rm M_{\rm SFR}$ is the star formed mass, $\rm t_{\rm d}$ is the delay time between the formation of the progenitor binary and the merging of the compact objects binary, $\rm Z$ is the metallicity and $\rm V$ the comoving cosmological volume.

The first term in the integral $\rm dN/dM_{\rm SFR}\,d\mathcal{M}\,dt_{\rm d}$ is related to stellar and binary evolution and represents the number of merging double compact objects (DCOs) per unit of star forming mass per bin of chirp mass and time delay. It can be evaluated via stellar and binary evolution simulations (see e.g. Dominik et al.  2012,  2015;  de Mink et al.  2013;  de Mink \& Belczynski 2015;  Belczynskiet al. 2016;  Spera \& Mapelli 2017;  Giacobbo \& Mapelli 2018;  Mapelli \& Giacobbo 2018; Chruslinska  et  al. 2018;  Spera et  al. 2019; Santoliquido et al. 2020). Various processes involved in stellar and binary evolution depend on metallicity 
(e.g. radiation-driven stellar wind mass loss rates, core-collapse physics, mass transfer characteristics and stability) and so the number of merging BH/NS binaries that form per unit mass formed in stars also varies with this quantity.

The second term $\rm d\dot M_{\rm SFR}/dV\,dZ$ is instead related to galaxy evolution: it represents the star forming mass per units of time, comoving volume and metallicity, i.e. it is the star formation rate (SFR) density per metallicity bin. There are two main ways to estimate it: exploiting the results of cosmological simulations (e.g., Mapelli et al. 2017; O’Shaughnessy et al. 2017; Lamberts et al. 2018; Mapelli \& Giacobbo 2018; Artale et al. 2019) or using empirical recipes concerning the cosmic SFR density and metallicity distributions inferred from observations (e.g., Belczynski et al. 2016; Lamberts et al. 2016; Cao et al. 2018; Elbert et al. 2018; Li et al. 2018; Boco et al. 2019; Chruslinska \& Nelemans 2019; Neijssel et al. 2019; Santoliquido et al. 2020). 

The main focus of this work is to revise the different empirical approaches pursued to compute the galactic term, trying to quantify the impact of different choices and to understand their advantages and drawbacks; moreover, we shall propose new ways to compute it. We stress that the methods discussed in the present paper to compute the galactic term are purely based on observations and on empirically derived scaling relations, and do not rely on semi analytical models or simulations. Finally, we study the effects of the different prescriptions on the merging rates and on the properties of merging binaries.

The plan of the paper is as follows. In section \ref{sec:galaxy_statistics} we compare the two main empirical ways to compute the cosmic SFR density: via the luminosity/SFR functions (SFRF) or via the galactic stellar mass functions (GSMF) and the main sequence (MS) of star forming galaxies. In section \ref{sec:metallicity} we present the two main scaling relations to empirically assign metallicity to galaxies (the Fundamental Metallicity Relation (FMR) or a Mass Metallicity Relation (MZR)) and we compute the galactic term $\rm d\dot M_{\rm SFR}/dV\,dZ$ combining the two possible metallicity scaling relations with the two galaxy statistics. In section \ref{sec:merging_rates}, we compute the merging rates and some properties of the compact binaries for these different prescriptions, basing on the outcomes of the STARTRACK binary evolution simulations as to compute the stellar factor $\rm dN/dM_{\rm SFR}dM_{\bullet\bullet}dt_{\rm d}$. Finally, in section \ref{sec:conclusions} we summarize our main findings.

Throughout this work, we rely on the standard flat $\Lambda\rm CDM$ cosmology (Planck Collaboration 2019) with cosmological parameters: $\Omega_{\rm M}=0.32$, $\Omega_{\rm b}=0.05$, $\rm H_0=67\,km\,s^{-1}\,Mpc^{-1}$. The Chabrier IMF (2003, 2005; see also Mo et al. 2010) is adopted, with mass range $0.08-150\,\rm M_\odot$. A value $Z_\odot=0.0153$ for the solar metallicity and $12+\log(\rm O/H)_\odot=8.76$ for the solar oxygen abundance is adopted (Caffau et al. 2011).

\section{Cosmic star formation rate and galaxy statistics}\label{sec:galaxy_statistics}
The first important ingredient in the computation of the factor $\rm d\dot{M}_{\rm SFR}/dVdZ$ is constituted by the cosmic SFR density $\rm d\dot{M}_{\rm SFR}/dV$, representing the average rate at which new stars are formed in the Universe at different redshifts per unit comoving volume. There are mainly two different ways to compute and exploit it, that are recalled and discussed below.

\subsection{SFR/Luminosity functions}\label{subsec:SFR_functions}

The most direct approach to compute the cosmic SFR density relies on the galaxies star formation rate functions (SFRF) $\rm dN/dV\,d\log\psi$ at different redshifts, representing the number density of galaxies per logarithmic bin of SFR ($\psi$). The SFRF can be computed from the UV and IR luminosity functions of galaxies (see Mancuso et al. 2016a; Boco et al. 2019), since luminosity can be converted into SFR (e.g., Kennicutt 1998; Kennicutt \& Evans 2012). 

In fact, the SFR of a galaxy can be related to its UV luminosity which mainly comes from young, blue stars. However, since dust absorbs UV radiation and re-emits it in the mid and far-IR band, the SFR estimated only from UV luminosities can be substantially underestimated. Nevertheless it is still possible, for galaxies with a relatively low SFR ($\psi\lesssim 30-50\,\rm M_\odot/yr$) and relatively small dust content, to estimate it from UV data alone using standard UV slope corrections (see Meurer et al. 1999; Calzetti et al. 2000; Bouwens et al. 2015). Therefore, the SFRF for $\psi\lesssim 30-50\,\rm M_\odot/yr$ can be well constrained using data from deep UV surveys (see Wyder et al. 2005; Oesch et al. 2010; van der Burg et al. 2010; Cucciati et al. 2012; Finkelstein et al. 2015; Alavi et al. 2016; Bouwens et al. 2016, 2017; Bhatawdekar et al. 2018). Contrariwise, for what concerns highly star-forming galaxies with $\psi\gtrsim 30-50\,\rm M_\odot/yr$ that are much more rich in dust, UV corrections tend to fail (see Silva et al. 1998; Efstathiou et al. 2000; Coppin et al. 2015; Reddy et al. 2015; Fudamoto et al. 2017); as a consequence, the estimates of the SFR must be based on far-IR/(sub)mm wide-area surveys (see Lapi et al. 2011; Gruppioni et al. 2013, 2015; Gruppioni \& Pozzi 2019; Magnelli et al. 2013). However, given the sensitivity limit of far-IR surveys, the shape of the SFR functions at the bright end becomes progressively uncertain at $z\gtrsim 3$. Still relevant constraints in this regime have been obtained from deep radio surveys (Novak et al. 2017), from far-IR/(sub)millimeter stacking (see Rowan-Robinson et al. 2016; Dunlop et al. 2017) and super-deblending techniques (see Liu et al. 2018), and from targeted far-IR/(sub)millimeter observations of significant yet not complete samples of star-forming galaxies (e.g., Riechers et al. 2017; Marrone et al. 2018; Zavala et al. 2018) and quasar hosts (e.g., Venemans et al. 2017, 2018; Stacey et al. 2018). Moreover, very recently, Gruppioni et al. (2020) estimated the total IR luminosity functions up to redshift $z\lesssim 6$ from a sample of $56$ galaxies serendipitously detected by ALMA in the COSMOS and ECDFS fields, finding a pleasant agreement with previous far-IR/(sub)mm Hershel data, within the still large observational uncertainties.

The aforementioned set of data can be fitted via a simple Schechter function:
\begin{equation}
    \frac{\rm dN}{\rm d\log\psi\,dV}(\psi, t)=\mathcal{N}(z)\left[\frac{\psi}{\psi_{\rm c}(z)}\right]^{1-\alpha(z)}\rm e^{-\psi/\psi_{\rm c}(z)}
    \label{eq:dNdlogsfr_lapi}
\end{equation}
in terms of three fitting parameters: $\mathcal{N}(z)$, $\psi_{\rm c}(z)$ and $\alpha(z)$ (see Table 1 of Mancuso et al. 2016a); in Fig. \ref{fig:sfr_functions}, left panel, we show the datasets mentioned above and the fitted SFRF. At $z\gtrsim 1$ heavily obscured, strongly star forming galaxies populate the bright end of the SFR functions; these galaxies are the progenitors of local massive ellipticals (ETGs) with final stellar mass $M_\star\gtrsim$ a few $\times 10^{10}M_\odot$. Mildly star forming objects, instead, populate the faint end and will end up in spheroid-like objects with rather low stellar mass ($\lesssim 10^{10}M_\odot$). Finally, late type disk galaxies (LTGs), with SFR of a few solar masses per year, are well traced by the UV-inferred SFR function at $z\lesssim 1$. This is confirmed also by the link between the SFRF and the stellar mass function of different morphological types, obtained via a continuity Eq.approach by Lapi et al. (2017).

From the SFRF, the cosmic SFR density can be easily estimated as:
\begin{equation}
    \frac{\rm d\dot{M}}{\rm dV}(\rm t)=\int\rm d\log\psi\,\psi\,\frac{\rm dN}{\rm d\log\psi\,dV}(\psi, t)
    \label{eq:dotMdV_lapi}
\end{equation}

The resulting determination of the cosmic SFR density is shown as a dot-dashed black line in Fig. \ref{fig:cosmic_sfrd}. It can be noticed that the cosmic SFR computed in this way tends to be appreciably higher with respect to most of the previous determinations. This is due to the recent discovery via IR and far-IR/submm observations with Hershel and ALMA, of a significant number of dusty star forming galaxies very attenuated or even invisible in the optical/UV bands. Such dusty galaxies, featuring an extremely high level of star formation ($\sim 50- 3000\,\rm M_\odot/yr$), seem to have a significant impact on the total star formation at $2\leq z\leq 6$ (see e.g. Wang et al. 2019; Gruppioni et al. 2020; Smail et al. 2020). Therefore, since our SFRF fit is based also on data coming from recent far-IR/(sub)mm surveys, the resulting cosmic SFR density is larger and more in agreement with the recent IR data (see e.g. Casey et al. 2018; Rowan-Robinson et al. 2016).

\subsection{Stellar mass functions + main sequence}\label{subsec:mass_functions}

Another method to estimate the cosmic SFR density is convolving the stellar mass functions of star forming galaxies at different redshifts with a distribution around the main sequence. This approach has been adopted in Chruslinska \& Nelemans (2019); here we recall it and improve it by adding a simple treatment of starburst galaxies.

\begin{figure*}
    \centering
    \includegraphics[width=0.47\textwidth]{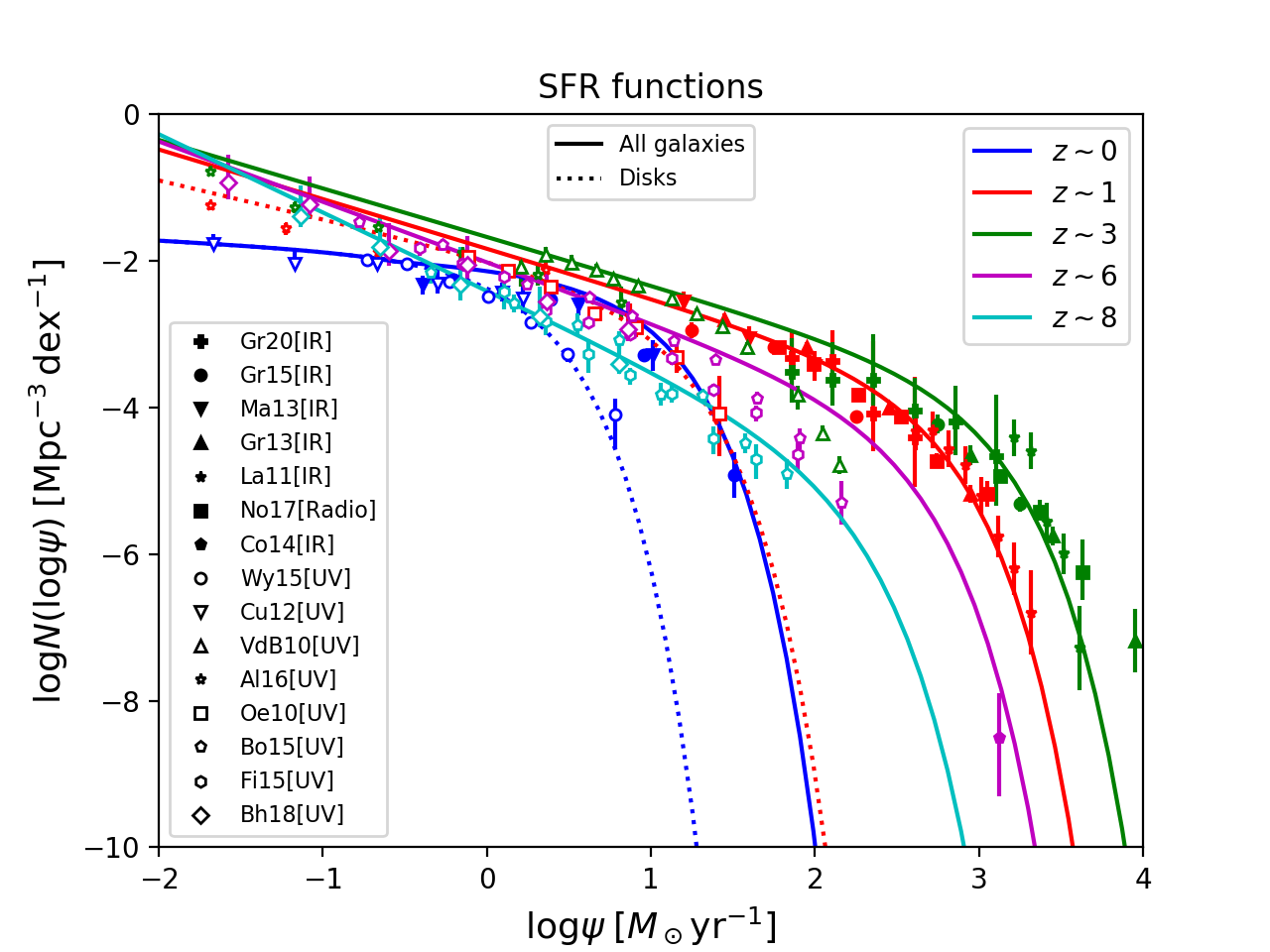}
    \includegraphics[width=0.47\textwidth]{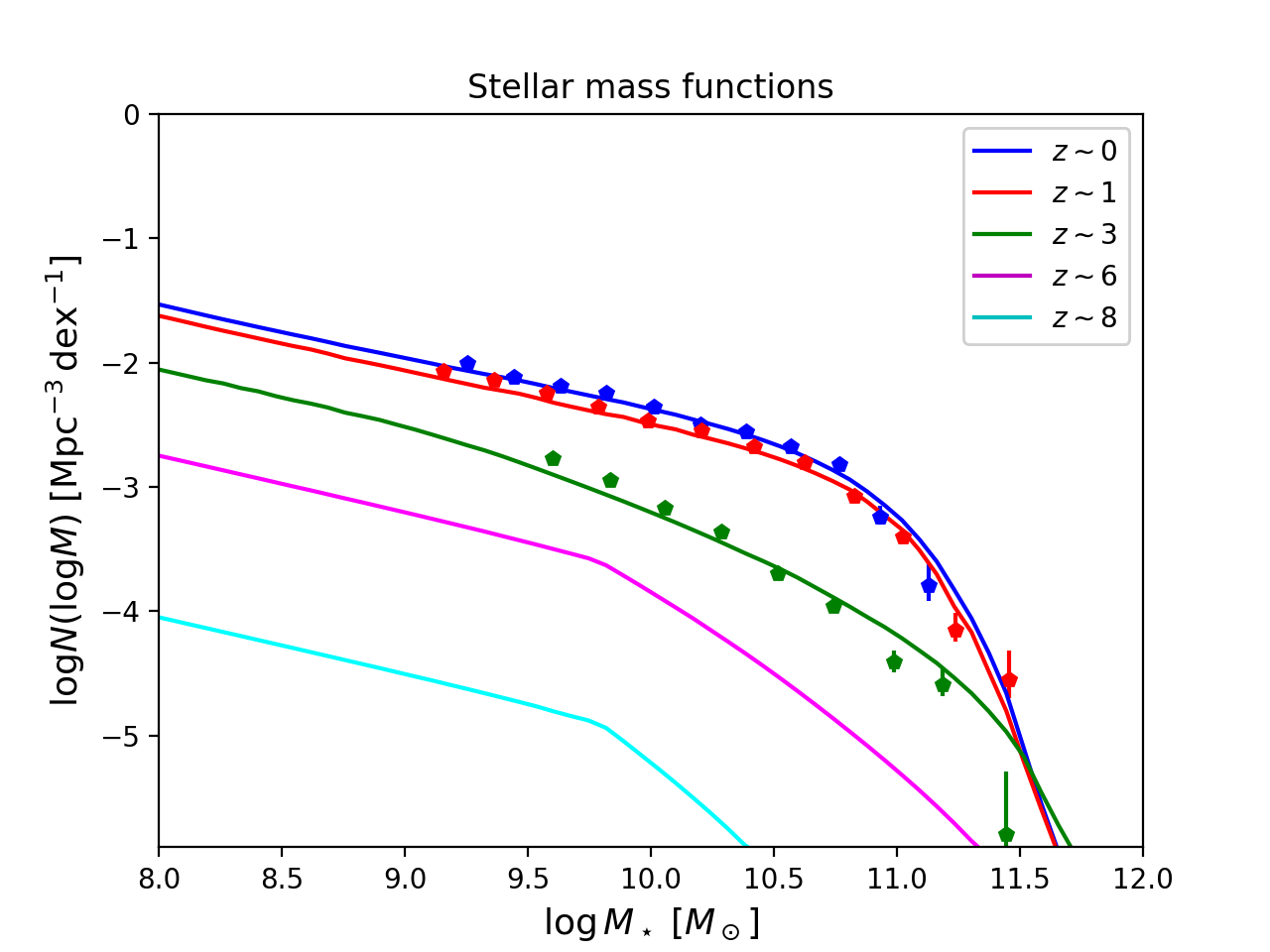}
    \caption{Left panel: SFR functions at redshifts $z=0$ (blue), 1 (red), 3 (green), 6 (magenta), and 8 (cyan). Solid lines show the rendition from UV plus far-IR/(sub)millimeter/radio data, referring to the overall population of galaxies; dotted lines (only plotted at $z\simeq0$ and $1$) show the rendition from (dust- corrected) UV data, referring to disk galaxies. UV data (open symbols) are from van der Burg et al. (2010; triangles), Bouwens et al. (2016, 2017; pentagons), Finkelstein et al. (2015; hexagons), Cucciati et al. (2012; inverse triangles), Wyder et al. (2005; circles), Oesch et al. (2010; squares), Alavi et al. (2016; stars), Bhatawdekar et al. (2018; rhombus); far-IR/(sub)millimeter data (filled symbols) are from Gruppioni et al. (2020; filled plus),Gruppioni et al. (2015; circles), Magnelli et al. (2013; inverse triangles), Gruppioni et al. (2013; triangles), Lapi et al. (2011; stars), and Cooray et al. (2014; pentagons); radio data are from Novak et al. (2017; squares). Right panel: Stellar mass functions for star forming galaxies at redshifts $z=0$ (blue), 1 (red), 3 (green), 6 (magenta), and 8 (cyan). Data points are taken from Davidzon et al. 2017.}
    \label{fig:sfr_functions}
\end{figure*}

The galaxy stellar mass is routinely estimated via near-IR data and broadband SED fitting (e.g., da Cunha et al. 2008; Boquien et al. 2019). The star forming galaxies stellar mass function (GSMF) $\rm dN/dV\,d\log M_\star$ has been determined at different redshifts by several authors (e.g. Ilbert et al. 2013; Muzzin et al. 2013; Tomczak et al. 2014; Davidzon et al. 2017). A good review of the different determinations can be found in Chruslinska \& Nelemans (2019), where the authors provide an average fit between many different works. Here we adopt their fit, and in particular their prescription for a redshift independent slope at the faint end (see Fig. \ref{fig:sfr_functions}, right panel and Fig. 3 in Chruslinska \& Nelemans 2019, solid lines).

In order to compute the cosmic SFR density, the GSMF must be convolved with a distribution of SFR around the main sequence (MS) of star forming galaxies. The MS is a well known (approximately powerlaw) relation between the stellar mass of the galaxy and its SFR at a given redshift. It has been determined both observationally and theoretically in different works (see e.g. Daddi et al 2007; Rodighiero et al. 2011, 2015; Whitekar et al. 2014; Speagle et al. 2014; Schreiber et al. 2015; Mancuso et al. 2016b; Dunlop et al. 2017; Bisigello et al. 2018; Pantoni et al. 2019; Lapi et al. 2020). However, we point out that the MS shape and evolution with redshift is still debated, with relevant differences among various works; in particular, its behaviour at large masses is very uncertain, with some authors advocating a possible flattening (although it may be effectively due to contamination from passive galaxies). 

Note that the MS is only an average relation between mass and SFR; actually, star forming galaxies with fixed mass at a given redshift tend to be distributed in SFR following a double gaussian shape (see Sargent et al. 2012; Béthermin et al. 2012; Ilbert et al. 2015; Schreiber et al. 2015). This bimodal distribution highlights the existence of two kind of galaxy populations: the dominant population of main sequence galaxies (MSG), whose Gaussian distribution in SFR is centered around the MS value and the subdominant population of starburst galaxies (SBG), whose Gaussian distribution is centered around a SFR typically $\sim 3-4\sigma$ above the MS value. In the aforementioned works it is empirically found that the shape of the distribution is almost independent of the galaxy stellar mass and redshift. On the other hand, other recent studies, probing the SFR distribution of galaxies around the MS in a more extended range of masses and redshifts, found an increase of the starbursts fraction at low masses $\rm M_\star\leq 10^9M_\odot$ or at high redshifts $\rm z\geq2-3$ (see Caputi et al. 2017; Bisigello et al. 2018) . 

For the sake of simplicity, in the present work we describe the galaxy distribution in SFR at fixed mass and redshift via a double Gaussian shape with the same parameters indicated by Sargent et al. (2012), and a fixed starbursts fraction in each redshift and mass bins:
\begin{equation}
\begin{split}
    \frac{\rm dp}{\rm d\log\psi}(\psi|z,M_\star)&=A_{\rm MS}\exp{\left[-\frac{(\log\psi-\langle\log\psi\rangle_{\rm MS})^2}{2\sigma_{\rm MS}^2}\right]}+\\
    &+A_{\rm SB}\exp{\left[-\frac{(\log\psi-\langle\log\psi\rangle_{\rm SB})^2}{2\sigma_{\rm SB}^2}\right]}
    \label{eq:dpdlogpsi}
\end{split}
\end{equation}
where $A_{\rm MS}=0.97$ is the fraction of MSG,  $A_{\rm SB}=0.03$ the fraction of SBG, $\langle\log\psi\rangle_{\rm MS}$ the value given by the MS and representing the central value for the first Gaussian, $\langle\log\psi\rangle_{\rm SB}=\langle\log\psi\rangle_{\rm MS}+0.59$ the central value of the second Gaussian, $\sigma_{\rm MS}=0.188$ the one-sigma dispersion of the first Gaussian and $\sigma_{\rm SB}=0.243$ the dispersion of the starburst population.

Once the distribution in Eq. (\ref{eq:dpdlogpsi}) is convolved with the GSMF, one can reconstruct the SFRF of galaxies as
\begin{equation}
\begin{split}
    \frac{\rm dN_{\rm GSMF+MS}}{\rm d\log\psi dV}(z,\log\psi)=\int&\rm d\log M_\star\frac{\rm dN}{\rm d\log M_\star dV}(z,\log M_\star)\times\\
    &\times\frac{\rm dp}{\rm d\log\psi}(\log\psi|z,M_\star)
    \label{eq:dotNdVdpsi_reconstructed}
\end{split}
\end{equation}
In Sargent et al. (2012) and Ilbert et al. (2015) it is also demonstrated that such a convolution yields a good reconstruction of the luminosity functions. 

\begin{figure*}
    \centering
    \includegraphics[width=0.75\textwidth]{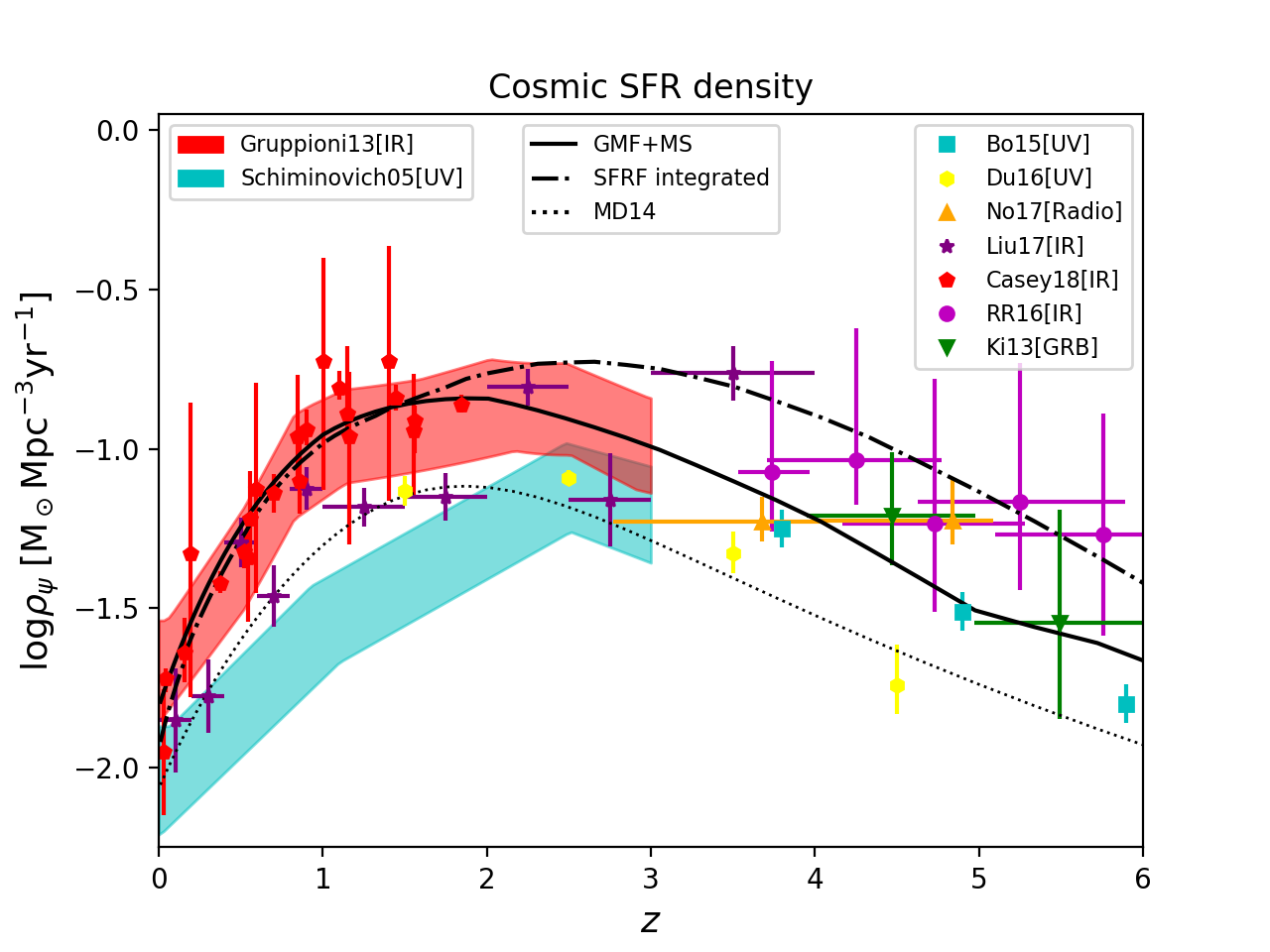}
    \caption{Cosmic SFR density as a function of redshift. The black solid line shows the result obtained integrating the SFR functions reconstructed from the stellar mass functions plus the main sequence (Eq.\eqref{eq:dotNdVdpsi_reconstructed}). The black dot dashed line shows, instead, the result of the integration of the SFR functions (Eq.\eqref{eq:dNdlogsfr_lapi}). For reference, the dotted line illustrates the determination by Madau \& Dickinson (2014). Data are from (dust-corrected) UV observations by Schiminovich et al. (2005; cyan shaded area) and Bouwens et al. (2015; cyan squares); ALMA submillimeter observations of UV-selected galaxies on the HUDF by Dunlop et al. (2017); VLA radio observations on the COSMOS field by Novak et al. (2017); Herschel far-IR observations by Gruppioni et al. (2013 red shaded area) and Casey (2018; red pentagons); Herschel far-IR stacking by Rowan-Robinson (2016; magenta circles); far-IR/(sub)millimeter observations from super-deblended data on the GOODS field by Liu et al. (2018); and estimates from long GRB rates by Kistler et al. (2009, 2013; green reversed triangles).}
    \label{fig:cosmic_sfrd}
\end{figure*}

Integrating the reconstructed SFRF in the Eq. above over the whole range of star formation, as in Eq.(\ref{eq:dotMdV_lapi}), we obtain the cosmic SFR density as a function of the cosmic time. 
In Fig. \ref{fig:cosmic_sfrd} we show the cosmic SFR density computed by integrating  the SFRF directly fitted from the data (as in Eq.\eqref{eq:dNdlogsfr_lapi}) as a dot-dashed line, and the one derived from the GSMF as in Eq.\eqref{eq:dotNdVdpsi_reconstructed} as a solid line. We find that the two determinations of the cosmic SFR density are in rather good agreement up to redshift $z\sim 2$. At $z>2$ the integration of the SFRF directly fitted from the data yields a larger cosmic SFR density, with the maximum differences being a factor $\sim 2.5$ at $z\sim 4.5$. These discrepancy, even if rather small, can be due to biases and selection effects arising respectively in the chosen determination of the SFR and stellar mass functions. For example, the shape of the faint end of the GSMF at high redshift is highly uncertain  and, using a mass function whose shape steepens toward higher redshifts (see e.g. Fig. 3, dashed lines in Chruslinska \& Nelemans 2019), drastically reduce the differences. Other factors that can produce these discrepancies are possible biases in the determination of the SFR from the UV+IR luminosity, in the shape of the main sequence or in the relative contributions of the main sequence and starburst populations. As for the latter, in Caputi et al. (2017) and Bisigello et al. (2018), it is pointed out that the population of starburst galaxies tends to increase at $z\gtrsim2$; keeping into account this trend can reduce the differences between the two cosmic SFR densities (Chruslinska et al. in preparation). All in all, from Fig. \ref{fig:cosmic_sfrd} we have shown that the two approaches yield a rather good agreement especially at $z\lesssim2$, and we have quantitatively characterized the differences toward higher redshifts. For reference, in Fig. \ref{fig:cosmic_sfrd} the classic determination of the cosmic SFR density by Madau \& Dickinson (2014) is also reported, which is seen to be a factor $\sim 2$ lower than some more recent IR data.

The main advantage of an approach based on the SFRF is that it is rather direct. Indeed the SFR is the main quantity we are interested in, since, provided an IMF, it gives the effective number of stars formed and so it provides a normalization for the DCOs merging rates. Starting from the SFRF we directly have a measure of the number density of galaxies with given SFR at a certain redshift. Instead, starting from the GSMF, the computation of the SFR requires a step more, since it involves the convolution with the main sequence and a correct modelization of the relative abundance of MS galaxies and starbursts.

On the other hand the GSMF provide a direct statistics of the star forming galaxies stellar masses and, as shown above, the distribution of SFRs at fixed stellar mass and redshift is well established in literature. Therefore, once the stellar mass is known, it is easy to associate a SFR and to use a scaling relation to infer its metallicity. Contrariwise, fixing the SFR and redshift, the association of a stellar mass is not straightforward from an empirical point of view, and some  assumptions about  the galaxy star formation  history (SFH) should be made, as we will see in subsection \ref{subsubsec:SFR+FMR}. It is therefore trickier to use a scaling relation to assign metallicity. Still, starting from the SFRF, it is possible to follow the chemical enrichment of a galaxy using a model of galaxy evolution, as done e.g. in Boco et al. (2019).

A final comparison concerns the possibility of disentangling different galactic populations using the two statistics mentioned. From the SFRF it can be determined the contribution to the total cosmic SFR density coming from late type disk galaxies (mainly traced by UV data) and progenitors of local early type galaxies (mainly traced by far-IR/(sub)mm data). From the GSMF, instead, it is possible to separate between main sequence galaxies and starburts. Understanding the contribution to the total SFR and the metallicities of different galactic populations can be important also for the association of a host galaxy to a GW event.

Another method commonly used in the literature to describe the SFR density per metallicity bin as a function of redshift is to combine one of the cosmic SFR determinations with a standalone metallicity distribution (see e.g. Belczynski et al. 2016; Cao et al. 2018; Li et al. 2018). In this approach, however, the link with the properties of star forming galaxies and their evolution is lost and it is not easy to retrieve an accurate cosmic metallicity distribution without passing through a galaxy statistics.

\section{Metallicity distribution}\label{sec:metallicity}
Along with the stellar mass and star formation rate, the metal content of the gas-phase of the ISM (i.e. the gas-phase metallicity, $Z_{\rm gas}$) is one of the key physical quantities that has to be considered in statistical galaxy evolution studies (for a review, see Maiolino \& Mannucci 2019). As it can be seen in Eq. (\ref{eq:merging rates}), it is a crucial ingredient also to compute the merging rates of DCOs, since many aspects of stellar and binary evolution depend on it. On global galaxy scales, the interplay between stellar mass, SFR and metallicity is naturally reflected by different scaling relations which encode informations on the galaxy evolutionary stage. There are different ways to parametrize $Z_{\rm gas}$ as a function of $M_{\star}$, redshift and/or SFR, either through a Mass Metallicity Relation ($\rm MZR$, e.g. Kewley \& Ellison 2008; Maiolino et al. 2008; Mannucci et al. 2009; Magnelli et al. 2012; Zahid et al. 2014; Genzel et al. 2015; Sanders et al. 2020a), or a Fundamental Metallicity Relation (FMR; e.g. Mannucci et al. 2010; Mannucci et al. 2011; Hunt et al. 2016; Curti et al. 2020).

The MZR is a correlation between $Z_{\rm gas}$ (typically measured from strong optical oxygen nebular emission lines as 12+$\log$(O/H)) and $M_{\star}$, and it is observationally found to be valid for objects with an $M_{\star}$ spanning over 5 orders of magnitude. In general, at fixed $M_\star$ the MZR predicts a decline in $Z_{\rm gas}$ towards higher redshifts and the level of redshift evolution is actively debated (Onodera et al. 2016; Sanders et al. 2020a). Some earlier works (e.g. Maiolino et al. 2008; Mannucci et al. 2009; Magnelli et al. 2012) found a slow evolution of the MZR out to $z\sim2$, but a very sharp decline in $Z_{\rm gas}$ of about 0.4-0.5 dex between $z=2.5$ and $z=3.5$, suggesting a huge drop in $Z_{\rm gas}$ in the early universe and creating somewhat tensions with the modern cosmological simulations of massive galaxy formation (e.g. Davé et al. 2017; Torrey et al. 2018). The problem of accurately determining $Z_{\rm gas}$ become strongly pronounced for high-$z$ ($z>3$) massive, dusty galaxies (see e.g. discussions in Tan et al. 2014; Liu et al. 2019; Tacconi et al. 2020), where the MZR should be extrapolated. Indeed, if a linear extrapolation of a sharply declining MZR is performed, very low values of metallicities ($12+\log(\rm O/H)<8.0$) are found at $z>3$ even for massive systems.

\begin{figure*}
    \centering
    \includegraphics[width=0.75\textwidth]{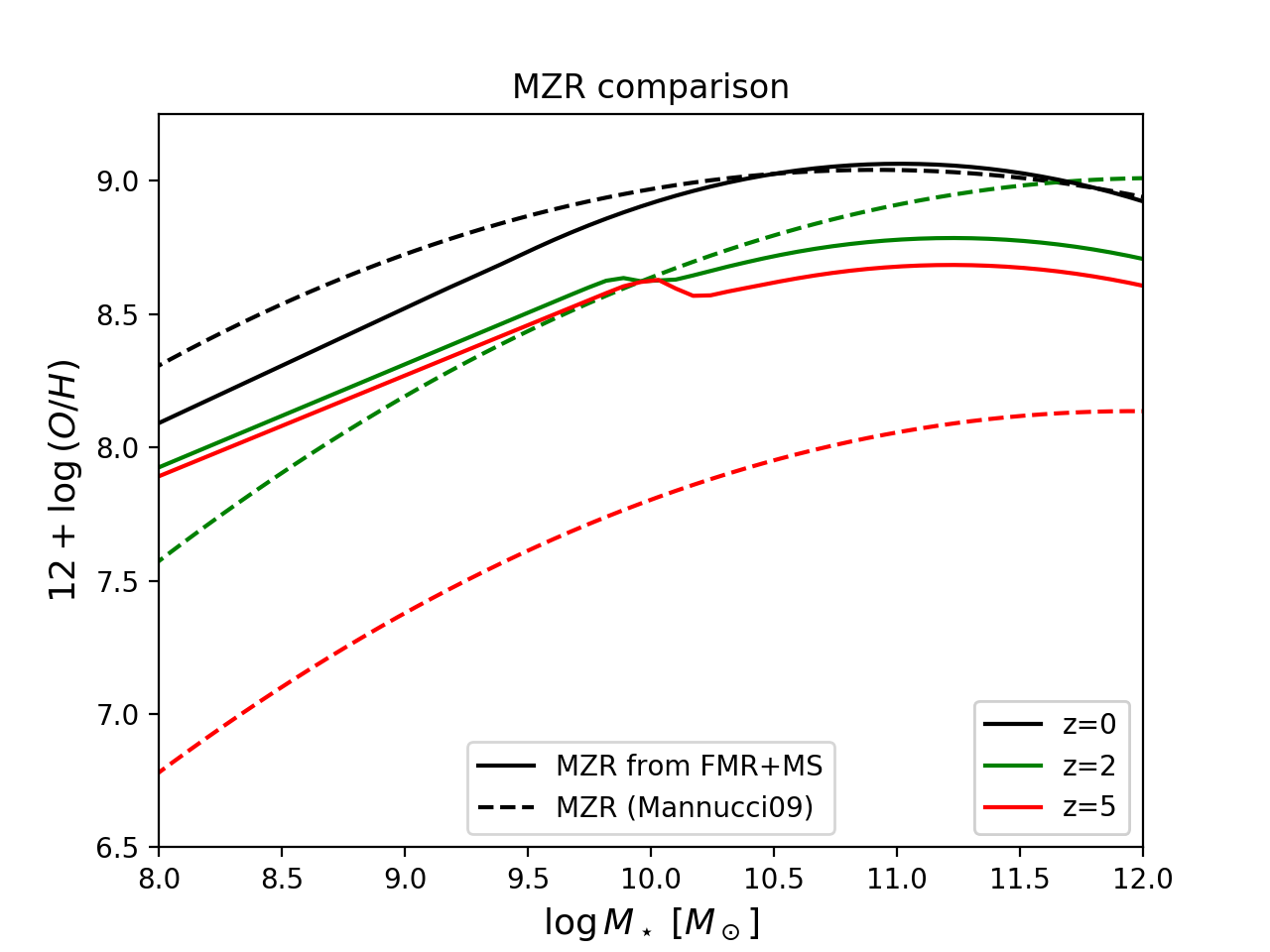}
    \caption{The average MZR relation $\rm \langle Z_{\rm MZR}\rangle$ computed convolving the FMR of Mannucci et al (2011) with the MS at different redshifts (solid lines), compared with the MZR determination of Mannucci et al. (2009) at different redshifts (dashed lines).}
    \label{fig:MZRFMR}
\end{figure*}

The FMR, instead, is a three parameter relation among $M_{\star}$, SFR, and $Z_{\rm gas}$. The inclusion of SFR is to account for the secondary dependence of metallicity on SFR initially observed in local SDSS galaxies (Mannucci et al. 2010) where $Z_{\rm gas}$ decreases increasing SFR at fixed $M_{\star}$. This dependence has been also confirmed over larger data sets: galaxies with the same stellar mass at the same redshift can have different metallicities due to their different SFR, showing a clear anti-correlation between $Z_{\rm gas}$ and sSFR (see e.g. Hunt et al. 2016). The FMR is thought to be almost redshift independent and this is confirmed by observations out to $z\sim 3.5$ (Mannucci et al. 2010; Hunt et al. 2016). Indeed, $z$ is not a parameter directly entering in the relation, and the metallicity evolution with redshift at fixed stellar mass can be traced back to the redshift evolution of the SFR (or sSFR), described by the main sequence. Therefore the extrapolation of the FMR at $z>3.5$ can be done following the redshift evolution of the main sequence, which is determined out to $z\sim 6$. This originates a rather shallow decline of metallicity with redshift.

Thus, while the level of redshift evolution for the FMR and MZR at $z\lesssim 2$ is somewhat comparable, the evolution of the two relations becomes completely different at $z\gtrsim 3$. To explicitly show these differences we put the two relations on the same ground computing a kind of averaged MZR from the FMR. We do this, at fixed stellar mass and redshift, averaging the metallicity given by the FMR over the distribution of SFRs around the MS in Eq.\eqref{eq:dpdlogpsi}:
\begin{equation}
    \langle \rm Z_{\rm MZR}\rangle(\rm z,M_\star)=\int\rm d\log\psi\frac{\rm dp}{\rm d\log\psi}(\psi|\rm z,M_\star)Z_{\rm FMR}(M_\star,\psi)
    \label{eq:MZR from FMR}
\end{equation}
In Fig. \ref{fig:MZRFMR} we show this averaged MZR (solid lines), computed from the Mannucci et al (2011) FMR which provides an updated version of the FMR presented in Mannucci et al (2010) for lower mass galaxies. In the Figure, for comparison, it is also reported the MZR of Mannucci et al. (2009) (dashed lines), linearly extrapolated at $z>3.5$. It can be noted that, while at low redshifts ($z\lesssim 2$) the two relations give comparable results, at high redshifts ($z\gtrsim3$) the MZR evolution is very rapid yielding values of metallicity much lower than those obtained from the FMR.

Trying to solve this tension is crucial, since, as seen in section \ref{sec:galaxy_statistics}, the amount of SFR is not negligible at $z>3$. However, the quest is rather challenging, since on the one hand, optical/near-IR spectroscopy suffers from large dust attenuation, on the other hand, the statistics of sources that have been spectroscopically studied through fine structure lines with ALMA is still limited (Boogard et al. 2019).

Nonetheless, very recent studies obtained a great agreement between the FMR and smoothly evolving MZR. Sanders et al. (2020) retrieved MOSDEF spectroscopy of a large sample of massive galaxies at $z>3$, and significantly improve the statistics upon past studies over the same redshift range. By employing a novel dust-correction method, they found a much slower evolution of MZR observing a very shallow metallicity decline of only 0.11 dex between $2.5<z<3.5$. This result greatly supports FMR and slowly evolving MZRs calibrated from UV+IR data (Genzel et al. 2015).

On top of that, an important evidence of a significant metal enrichment in the early Universe came from the novel dust mass ($M_{\rm dust}$) estimates in distant galaxies at $z>3-6$  (e.g.  da Cunha et al. 2015, Donevski et al. 2020; Ginolfi et al. 2020). For instance, by analysing a large sample of 300 massive ($M_{\star}>10^{10}\:M_{\odot}$) dusty galaxies in the COSMOS field observed with ALMA over a wide redshift range ($0.5<z<5.25$), Donevski et al. 2020 show that, in order to explain the observed $M_{\rm dust}$, their $Z_{\rm gas}$ are, on average, close to solar (12+log(O/H)=8.64 and 12+log(O/H)=8.52 for MS and SB galaxies, respectively). These values are in great agreement with recent direct $Z_{\rm gas}$ measurements through  $\rm [NII]\lambda6584/H\alpha$ ratio by Shapley et al. (2020) for dusty galaxies within the same mass range at $z\sim 2$.

All of that complements classical arguments from stellar archaeology suggesting a fast metal enrichment of galaxies even at high z. Indeed the study of stellar emission in local massive early type galaxies can place very good constraints on their metallicity evolution: stars observed in these galaxies, formed typically at high redshifts, are found to be almost coeval and $\alpha$-enhanced, indicating a short ($<1\,\rm Gyr$) burst of high star formation stopped by some form of energetic feedback (e.g., Romano et al. 2002; Thomas et al. 2005, 2010; Gallazzi et al. 2006; Johansson et al. 2012). Their stellar metallicity ranges from $0.5\,Z_\odot-2\,Z_\odot$ (see Thomas et al. 2010, Gallazzi et al. 2014; Maiolino \& Mannucci 2019) implying that their chemical enrichment should have been rather rapid. In Morishita et al. 2019 a sample of $24$ quiescent galaxies at $z\sim 2$ has been studied finding average stellar metallicities of $Z\sim 1.5-2\,Z_\odot$ (see also Saracco et al. 2020); in particular, the authors find that the relation between stellar mass and stellar metallicity of their sample shows no evolution with respect to the same relation for $z\sim 0$ galaxies in Gallazzi et al. (2014). Finally, a direct measure of metallicity through $\rm [O III]_{88\mu\rm m}/[N II]_{122\mu\rm m}$ line ratio in high redshift quasar hosts (up to $z\sim 7.5$) has been performed by several authors, showing solar and supersolar metallicity values with no signs of redshift evolution (see e.g. Juarez et al. 2009; Novak et al. 2019; Onoue et al. 2020; Li et al. 2020).

These numerous findings point towards the need of a rapid metal enrichment in the distant Universe, a scenario that has recently been proposed theoretically by several authors (Asano et al. 2013; Béthermin et al. 2015; Popping et al. 2017; Vijayan et al. 2019; Pantoni et al. 2019; Lapi et al. 2020). We have also explored predictions on the metallicity evolution from the state-of-the-art cosmological simulations (Davè et al. 2019) that self-consistently model gas and dust under standard IMF. By looking at different snapshots over the redshift range $0<z<5$ for the most massive objects ($10^{10}\,\rm M_\odot<M_\star<10^{11}\,\rm M_\odot$) we found very shallow $Z_{\rm gas}$ evolution of only $0.3\,\rm dex$ drop from $z\sim 0$ to $z\sim 5$. This further strengthen the above cited observational findings that can suffer of selection biases.

Evidence for substantial metal content is also found for less massive galaxies ($10^{9}<M_{\star}<10^{9.5}\,M_{\odot}$) in the epoch of re-ionization ($6<z<9$, Jones et al. 2020; Strait et al. 2020). These studies claimed that the observed $Z_{\rm gas}$ can be achieved by extrapolating FMR or slowly evolving MZRs.

All these reasons motivate us to apply prescriptions based on FMR as a main scaling relation to infer the metal properties of galaxies at high $z$'s (see the next Section). However, given the substantial uncertainties, in Section 3.2 we also show the case in which a MZR with a rapid decrease in $Z_{\rm gas}$ with redshift (e.g.,  Mannucci et al. 2009) is assumed as representative for the whole population of galaxies at $z>3$.

\subsection{The galactic term computed through a FMR}\label{subsec:metal_FMR}

Given all the arguments above we compute the galactic term $\rm d\dot{M}_{\rm SFR}/dV\,dZ$ using the FMR presented in Mannucci et al. (2011), assuming that we can extrapolate the FMR in the same form even at $z>3.5$, as said in section \ref{sec:metallicity}. Since the FMR is a relation between stellar mass, SFR and metallicity ($Z_{\rm FMR}=Z_{\rm FMR}(M_\star,\psi)$) we can use both the GSMF (subsection \ref{subsubsec:GSMF+FMR}) and the SFRF (subsection \ref{subsubsec:SFR+FMR}) as galaxy statistics to perform the computation.

\subsubsection{GSMF + FMR}\label{subsubsec:GSMF+FMR}

Fixing redshift and stellar mass, we can derive a distribution in SFR as in Eq.\eqref{eq:dpdlogpsi}; therefore the factor $\rm d\dot{M}_{\rm SFR}/dV\,dZ$ can be computed as:
\begin{equation}
\begin{split}
    \frac{\rm d\dot{M}_{\rm SFR}}{\rm dVdZ}(Z|z)=\int&\rm d\log M_\star\frac{\rm dN}{\rm dVd\log M_\star}(\log M_\star|z)\times\\
    &\begin{split}\times\int&\rm d\log\,\psi\,\psi\frac{\rm dp}{\rm d\log\psi}(\psi|z,M_\star)\times\\
    &\times\left.\frac{\rm dp}{\rm dZ}\right|_{\rm FMR}(Z|Z_{\rm FMR}(M_\star,\psi))
    \end{split}
    \end{split}
    \label{eq:sfrd_FMR}
\end{equation}
where $\rm \left.dp/d\log Z\right|_{\rm FMR}(Z|Z_{\rm FMR}(M_\star,\psi))\propto\exp{\left[-(\log Z-\log Z_{\rm FMR}(M_\star,\psi))^2/2\sigma^2_{\rm FMR}\right]}$ is just a log-normal distribution around the logarithmic metallicity value setted by the FMR at fixed stellar mass and SFR, the factor $\rm dN/dV\,d\log M_\star$ represents the GSMF, and the factor $\rm dp/d\log\psi$ is the distribution in SFR around the MS computed as in Eq.\eqref{eq:dpdlogpsi}. Notice that, using the FMR, there is not an explicit redshift dependence on the value of the metallicity; the redshift dependence enters only indirectly through the GSMF and the distribution of SFR $\rm dp/d\log\psi$ around the MS value.

\begin{figure*}
\centering
\includegraphics[width=0.75\textwidth]{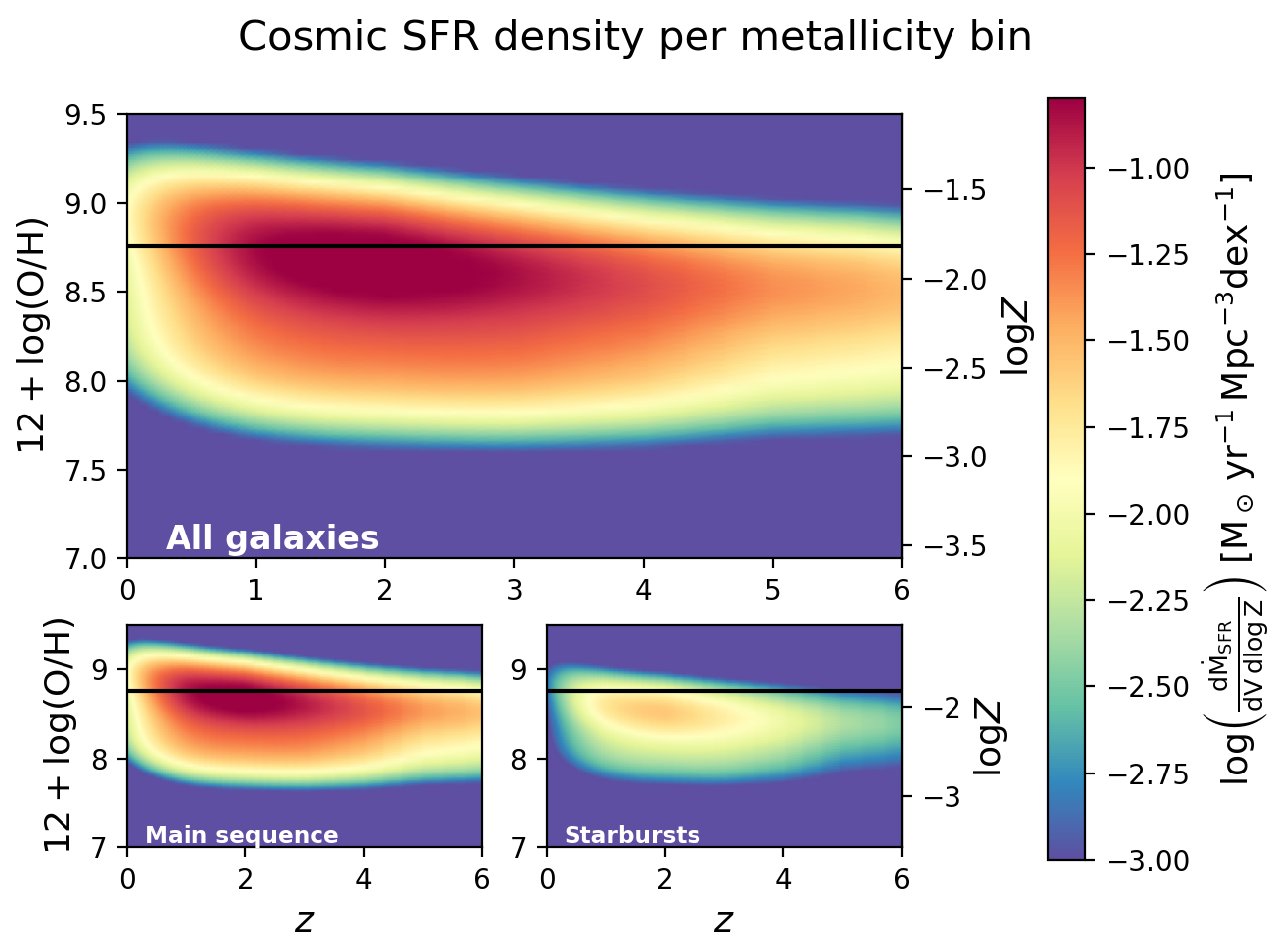}
\caption{Top panel: The factor $\log{(\rm d\dot{M}_{\rm SFR}/dV\,d\log{Z})}$ computed convolving the GSMF with the FMR of Mannucci et al. (2011) (color coded) as a function of redshift on the x axis and gas phase metallicity $12+\log{(\rm O/H)}$ on the left y axis; on the right y axis it is plotted the logarithm of the metallicity $\rm \log\,Z$ and the solar value is plotted as a black solid line.
Bottom left panel: The contribution to the cosmic SFR density coming from main sequence galaxies.
Bottom right panel: The contribution to the cosmic SFR density coming from starburst galaxies.}
\label{fig:GMF_FMRmannucci}
\end{figure*}

In Fig. \ref{fig:GMF_FMRmannucci} (top panel), we show the result of Eq.\eqref{eq:sfrd_FMR}, using the Mannucci et al. (2011) FMR and the GSMF from Chruslinska \& Nelemans (2019). The redshift dependence of the cosmic SFR density reflects the black solid line in Fig. \ref{fig:cosmic_sfrd}, as expected since the GSMF is used as starting point. As for the metallicity dependence, we notice that its redshift evolution is very mild: there is not a net evidence of a strong decrease with redshift of the metallicity at which star formation occurs, as expected looking at Fig. \ref{fig:MZRFMR}. In fact, while at $z\lesssim 2$ most of the star formation takes place at around solar values, at $z\sim4-5$ the typical values of metallicity at which star formation occurs are around $\rm Z\sim0.4-0.5\,Z_\odot$. The bottom left and bottom right panels show, respectively, the contribution of main sequence galaxies and starbursts. It can be noticed that the metallicity at which starbursts form stars tends to be slightly lower. This is natural, since, at fixed mass, the FMR predicts lower metallicities increasing the SFR.

\subsubsection{SFRF + FMR}\label{subsubsec:SFR+FMR}
The FMR can be used to assign metallicities even if the SFRF are chosen as galaxy statistics. The main difficulty is that, while at fixed redshift and stellar mass we are able to construct a distribution of SFRs, it is not clear how to derive
a distribution of galaxies stellar masses at fixed redshift and SFR; there are no works in literature facing the issue of deriving a stellar mass distribution from empirical data. This is why, in order to roughly estimate such stellar mass distribution, we must assume a star formation history for our galaxies. 

Since we are considering only star forming galaxies, the value of masses that they can assume at fixed redshift and SFR is less or equal than the value of mass given by the main sequence $\rm M_{\star,\rm MS}(z,\psi)$; all the values of mass larger than that represent quenched galaxies which are no more forming stars. Actually we will not sharply cut all the stellar masses above $\rm M_{\star,\rm MS}$, rather we put a Gaussian tail for masses $\rm M_\star\geq M_{\star,\rm MS}$. As for the mass distribution for stellar masses smaller than the MS mass ($\rm M_\star<M_{\star,\rm MS}$) we should make some assumptions on the galaxies SFH. 

For ETG progenitor galaxies, SED-modeling studies (e.g., Papovich et al. 2011; Smit et al. 2012; Moustakas et al. 2013; Steinhardt et al. 2014; Cassar\'a et al. 2016; Citro et al. 2016) suggest that the SFH can be described with a truncated power-law shape rising with a shallow slope $\leq 0.5$ for a quite short star formation timescale $\leq 1\rm Gyr$. On the other hand, late type disk dominated galaxies tend to have, on average, a SFH exponentially declining over rather long star formation timescales $\tau_\psi\sim$ several Gyr (see Chiappini et al. 1997; Courteau et al. 2014; Pezzulli \& Fraternali 2016; Grisoni et al. 2017; Lapi et al. 2020). Even if the star formation timescales are very different, the SFRs in both cases are nearly constant with time: for the ETG progenitors the SFR range is not larger than a factor $\sim 1.5$ for most of their lifetime, while for LTGs the SFR changes only of a factor $\sim 2.5$ over $\sim 8-9\,\rm Gyr$. For this reasons, for the sake of simplicity, we assume a constant SFH for the galaxies considered in this work. Under this assumption, the stellar mass of a galaxy increases linearly with time, with a slope set by its SFR. The logarithmic distribution of masses, at fixed redshift and SFR, is therefore proportional to the mass itself and can be written as:
\begin{equation}
\begin{split}
    &\frac{\rm dp}{\rm d\log M_\star}(\rm M_\star|z,\psi)\propto\\
    \propto&
    \begin{cases}
    \rm M_\star & M_\star<M_{\star,\rm MS}\\
    \rm M_{\star,MS}\times\exp{\left(-\frac{(\log{\rm M_\star}-\log{\rm M_{\star,\rm MS}})^2}{2\sigma_{\rm M_\star}}\right)} & M_\star\geq M_{\star,\rm MS}
    \end{cases}
    \end{split}
    \label{eq:dpdm}
\end{equation}
normalized to unity. Actually, this is a crude approximation of what has been done in Mancuso et al. (2016b), where the authors shows how to reproduce the MS with similar prescriptions.

Using the SFRF as a starting point and the distribution in Eq.\eqref{eq:dpdm} as stellar mass distribution at given $z$ and $\psi$, we can assign a metallicity to galaxies with the FMR and compute the factor $\rm d\dot{M}_{\rm SFR}/dVdZ$ as:
\begin{equation}
\begin{split}
    \frac{\rm d\dot{M}_{\rm SFR}}{\rm dV\,dZ}(Z|z)=\int&\rm d\log\psi\,\psi\,\frac{\rm dN}{\rm dV\,d\log\psi}\times\\
    &\begin{split}\times\int& \rm dM_\star\frac{\rm dp}{\rm dM_\star}(M_\star|z,\psi)\\
    &\times\left.\frac{\rm dp}{\rm dZ}\right|_{\rm FMR}(Z|Z_{FMR}(M_\star,\psi))
    \end{split}
\end{split}
    \label{eq:sfrd_FMR_SFR}
\end{equation}
where $\rm \left.dp/d\log Z\right|_{\rm FMR}(Z|Z_{\rm FMR}(M_\star,\psi))\propto\exp{\left[-(\log Z-\log Z_{\rm FMR}(M_\star,\psi))^2/2\sigma^2_{\rm FMR}\right]}$ is the same log-normal distribution around the central logarithmic value of metallicity set by the FMR appearing in Eq.\eqref{eq:sfrd_FMR}.

\begin{figure*}
\centering
\includegraphics[width=0.75\textwidth]{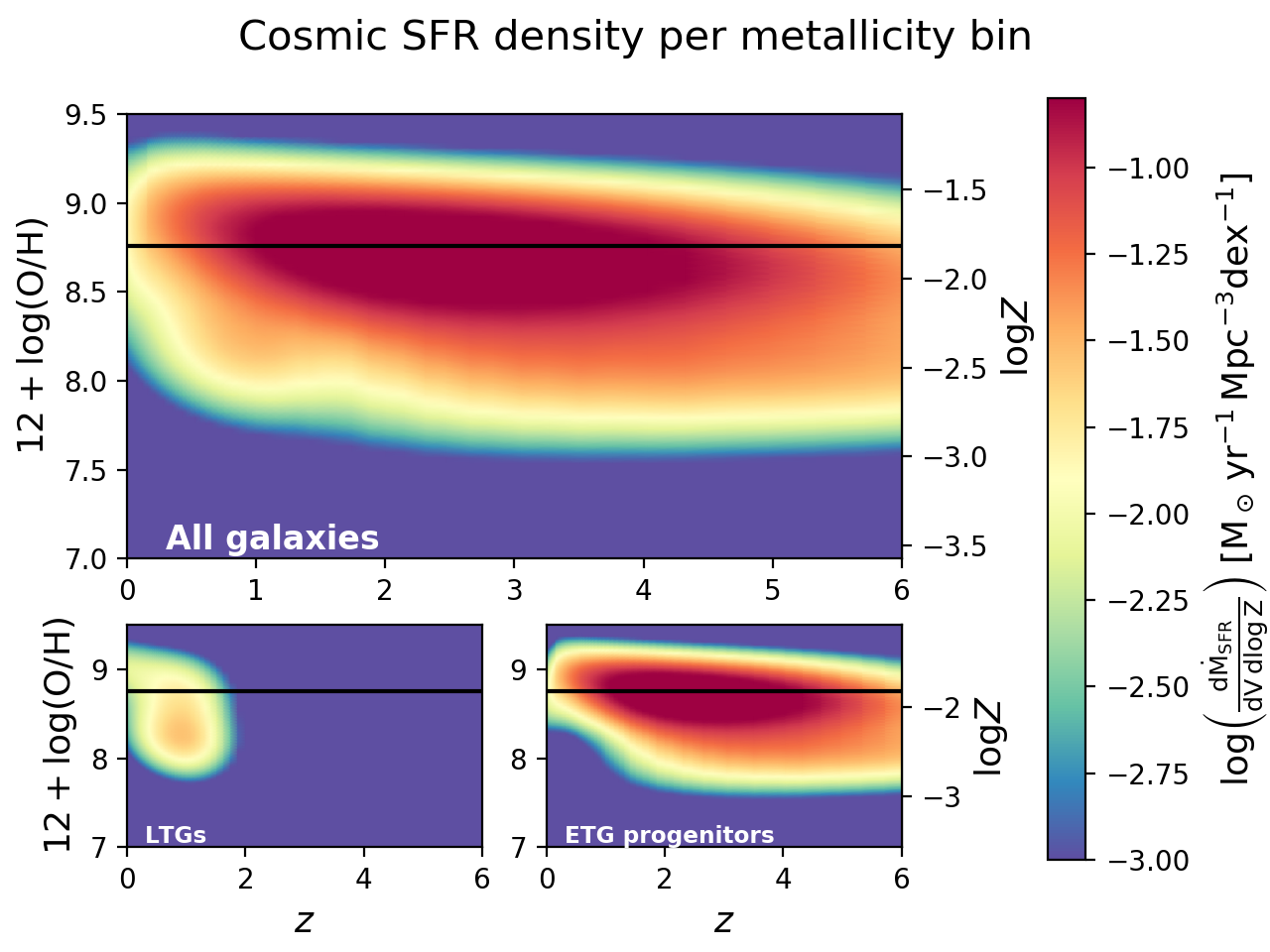}
\caption{Top panel: The factor $\log{(\rm d\dot{M}_{\rm SFR}/dV\,d\log{Z})}$ computed convolving the SFRF with the FMR of Mannucci et al. (2011) (color coded) as a function of redshift on the x axis and gas phase metallicity $12+\log{(\rm O/H)}$ on the left y axis; on the right y axis it is plotted the logarithm of the metallicity $\rm \log\,Z$ and the solar value is plotted as a black solid line.
Bottom left panel: Contribution to the cosmic SFR density coming from LTGs. 
Bottom right panel: Contribution to the cosmic SFR density coming from ETG progenitors.}
\label{fig:SFR_FMR}
\end{figure*}

The result of the computation in Eq.\eqref{eq:sfrd_FMR_SFR} is shown in Fig. \ref{fig:SFR_FMR}. It can be noticed that the redshift dependence of the SFR density reflects the shape of the cosmic SFR density derived by the integration of the SFRF (dot-dashed lines in Fig. \ref{fig:cosmic_sfrd}), with a broader peak slightly shifted towards $z\sim 2.5-3$, as expected, since the employed galaxy statistics is the same. The metallicity dependence on redshift is similar to Fig. \ref{fig:GMF_FMRmannucci}, since they share the same prescription to assign metallicity (the FMR). In particular the redshift decrease is mild also in this case, with most of the star formation occurring at solar metallicities for $z\lesssim 2$ and at $Z\sim 0.4-0.5\,Z_\odot$ for $z\sim 4-5$. In case the SFRFs are used as a starting point, it is more difficult to disentangle the contribution of main sequence galaxies with respect to starbursts. On the other hand, it is easier to look at the contribution to the total SFR density given by LTGs and ETG progenitors, as explained in section \ref{sec:galaxy_statistics}. These contributions are shown, respectively, in the bottom left and bottom right panels of Fig. \ref{fig:SFR_FMR}. It can be noticed that LTGs, having on average a lower stellar mass, tend to produce a tail of lower metallicity star formation even at low redshift. However, the bulk of the SFR density occurs at $z\gtrsim 1$ and it is given by ETG progenitors.

\begin{figure*}
    \centering
    \includegraphics[width=0.75\textwidth]{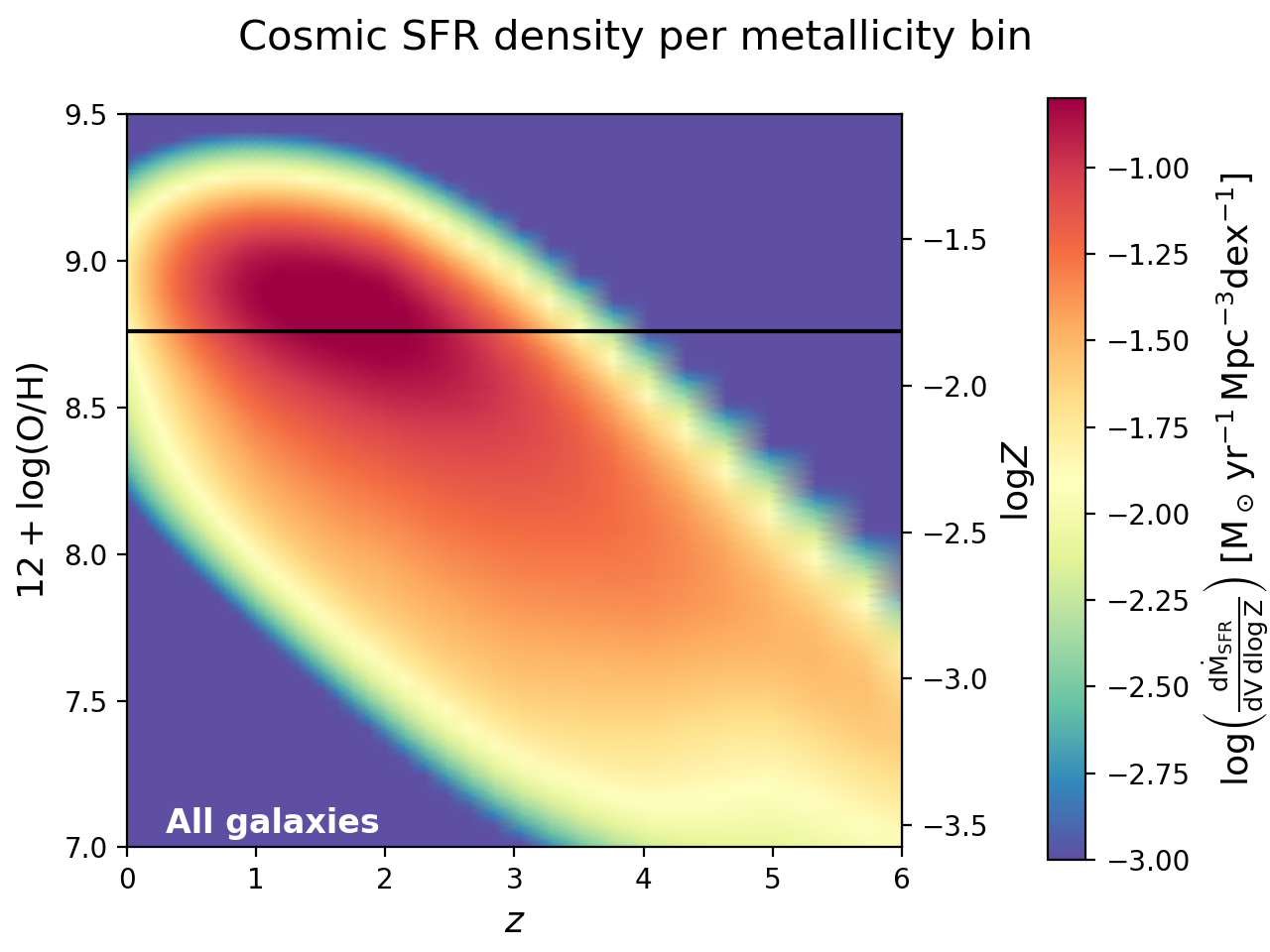}
    \caption{The factor $\log{(\rm d\dot{M_{\rm SFR}}/dV\,d\log{Z})}$ computed convolving the GSMF with the MZR of Mannucci et al. (2009) (color coded) as a function of redshift on the x axis and gas phase metallicity $12+\log{(\rm O/H)}$ on the left y axis; on the right y axis it is plotted the logarithm of the metallicity $\rm \log\,Z$ and the solar value is plotted as a black solid line.}
    \label{fig:GMF_MZR}
\end{figure*}

\subsection{The galactic term computed through a MZR}\label{subsec:metal_MZR}

In subsection \ref{subsec:metal_FMR} we computed the galactic term using a FMR which, as shown in our results, imply a shallow descrease of metallicity with redshift. We now compute the same factor $\rm d\dot{M}_{\rm SFR}/dV\,dZ$ using instead a sharply declining MZR (Mannucci et al. 2009), linearly extrapolated at $z>3.5$ to see how much this choice will impact on the final results.

The factor $\rm d\dot{M}_{\rm SFR}/dV\,dZ$ can be computed convolving the MZR with the stellar GSMF at a given redshift and using the distribution around the main sequence to assign a SFR to a galaxy with given stellar mass and redshift:
\begin{equation}
\begin{split}
    \frac{\rm d\dot{M}_{\rm SFR}}{\rm dVdZ}(Z|z)=\int&\rm d\log M_\star\frac{\rm dN}{\rm dV\,d\log M_\star}(\log M_\star|z)\times\\
    &\begin{split}&\times\left.\frac{\rm dp}{\rm dZ}\right|_{\rm MZR}(Z|Z_{\rm MZR}(z,M_\star))\times\\ 
    &\times\int\rm d\log\psi\frac{\rm dp}{\rm d\log\psi}(\psi|z,M_\star)\,\psi
    \end{split}
    \end{split}
    \label{eq:sfrd_MZR}
\end{equation}
where $\rm \left.dp/d\log Z\right|_{\rm MZR}(Z|Z_{\rm MZR}\,(z,M_\star))\propto\exp{\left[-(\log Z-\log Z_{\rm MZR}(z,M_\star))^2/2\sigma^2_{\rm MZR}\right]}$ is a log-normal distribution around the logarithmic value given by the MZR ($\log Z_{\rm MZR}$).

In Fig. \ref{fig:GMF_MZR} we show the resulting $\rm d\dot{M}_{\rm SFR}/\rm dV\,dZ$ (color code) as a function of redshift and metallicity. As for the redshift dependence of the cosmic SFR, it reflects the shape presented in Fig. \ref{fig:cosmic_sfrd} (solid lines) obtained using the GSMF as galaxy statistic, with a peak of star formation around $z\sim 2$. As for the metallicity dependence, at lower redshifts $z\lesssim2$ the metallicity stays rather high, similarly to the FMR cases, with most of the star formation occurring at slightly supersolar values, while at higher $z$ the metallicity starts to decline rapidly, with most of the star formation occurring at $\rm Z\leq 0.1\,Z_\odot$ at $z\gtrsim 4$, in contrast to the FMR cases in which the metallicity stays around $\sim 0.4\,\rm Z_\odot$, reaching values of $\sim 0.1\,Z_\odot$ only in the less massive systems. We stress that the differences between the two approaches are rather small at low redshifts $z\leq 2.5$ and start to be significant at higher redshifts, mainly in the regions where both the relations have been extrapolated. The main message here is that extrapolating the FMR, which is a redshift independent relation, can yield higher metallicity values with respect to a sharply declining MZR, more in agreement with the arguments discussed at the beginning of this section.
 
\section{Merging rates of compact binaries}\label{sec:merging_rates}

In this Section we show how the different galactic terms $\rm d\dot{M}_{\rm SFR}/dV\,dZ$ computed above impact on the merging rates and on the properties of merging compact binaries. The three cases for which the galactic term has been computed are: the stellar mass functions as galaxy statistics and the FMR to assign metallicity (GSMF+FMR case, Eq.\eqref{eq:sfrd_FMR}), the SFRF as statistics and the FMR to assign metallicity (SFRF+FMR, Eq.\eqref{eq:sfrd_FMR_SFR}), the stellar mass function as statistics and the MZR to assign metallicity (GSMF+MZR, Eq.\eqref{eq:sfrd_MZR}).

However, to compute the merging rates (Eq.\eqref{eq:merging rates}) we need not only the galactic term $\rm d\dot{M}_{\rm SFR}/dV\,dZ$ discussed throughout the paper, but also the term $\rm dN/dM_{\rm SFR}d\mathcal{M}dt_d$, depending on stellar and binary evolution, counting the number of merging events per units of star formed mass, chirp mass and time delay. In subsection \ref{subsec:stellar term} we describe the choices done to compute this term and then, in subsection \ref{subsec:merging rates}, we derive the merging rates for the three types of compact binaries: BH-BH, NS-NS and BH-NS. Finally, in subsection \ref{subsec:time delay}, we discuss about the time delay between the formation of the binary and the merger which will give us information on the typical ages of the stellar population of the galaxy hosting the merging event.

A general caveat for this Section is that, while we compare the results arising from the usage of different galactic terms $\rm d\dot{M}_{\rm SFR}/dV\,dZ$, the merging rates could even be strongly affected by the modelization of the factor $\rm dN/dM_{\rm SFR}d\mathcal{M}dt_{\rm d}$, as we will see in subsection \ref{subsec:stellar term}. So, the presented merging rates should not be intended as an exact determination, but just as the results of the different galactic prescriptions applied to a specific reference case for the stellar term $\rm dN/dM_{\rm SFR}d\mathcal{M}dt_{\rm d}$.

\begin{figure*}
    \centering
    \includegraphics[width=0.5\textwidth]{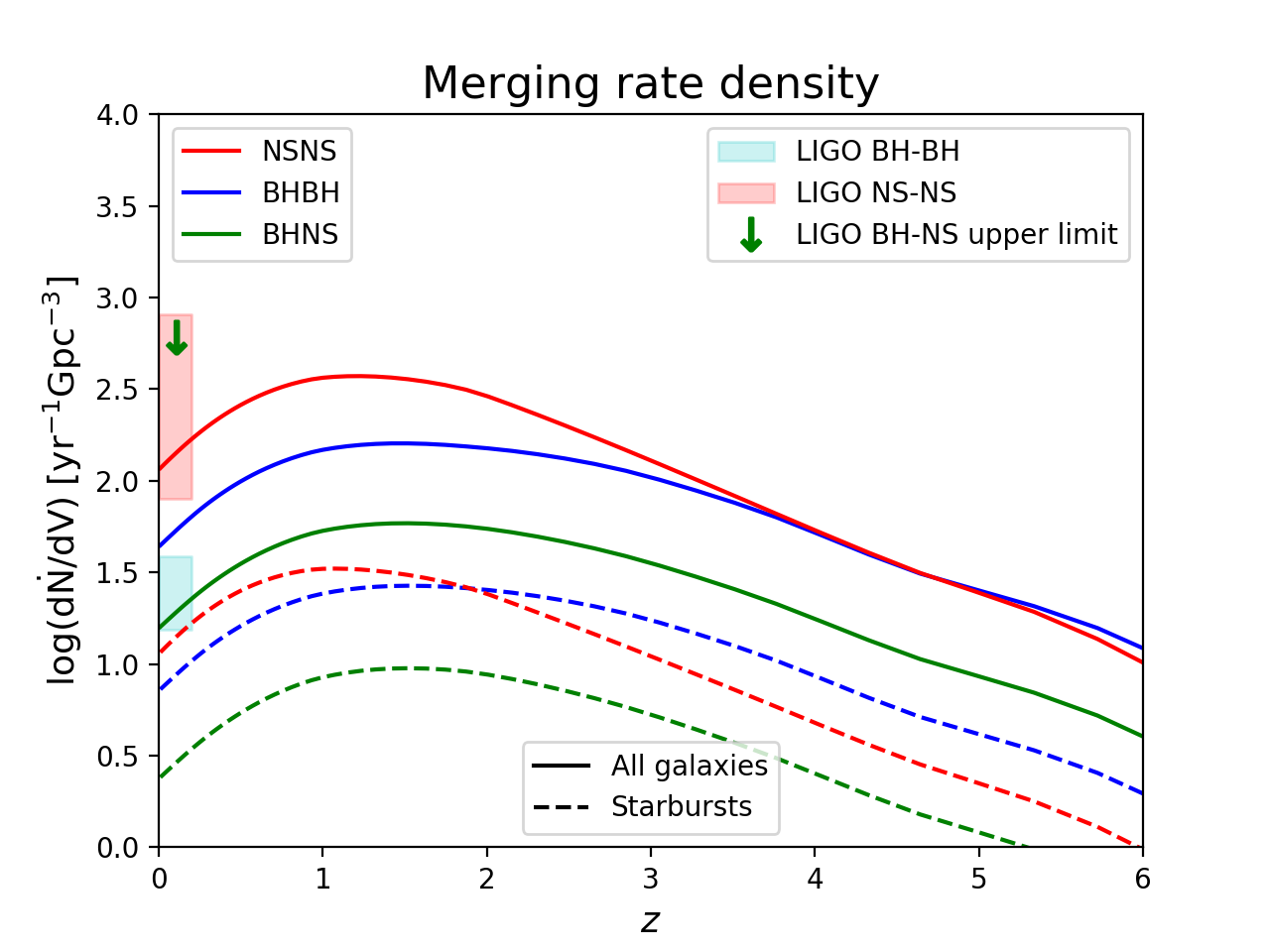}
    \includegraphics[width=0.5\textwidth]{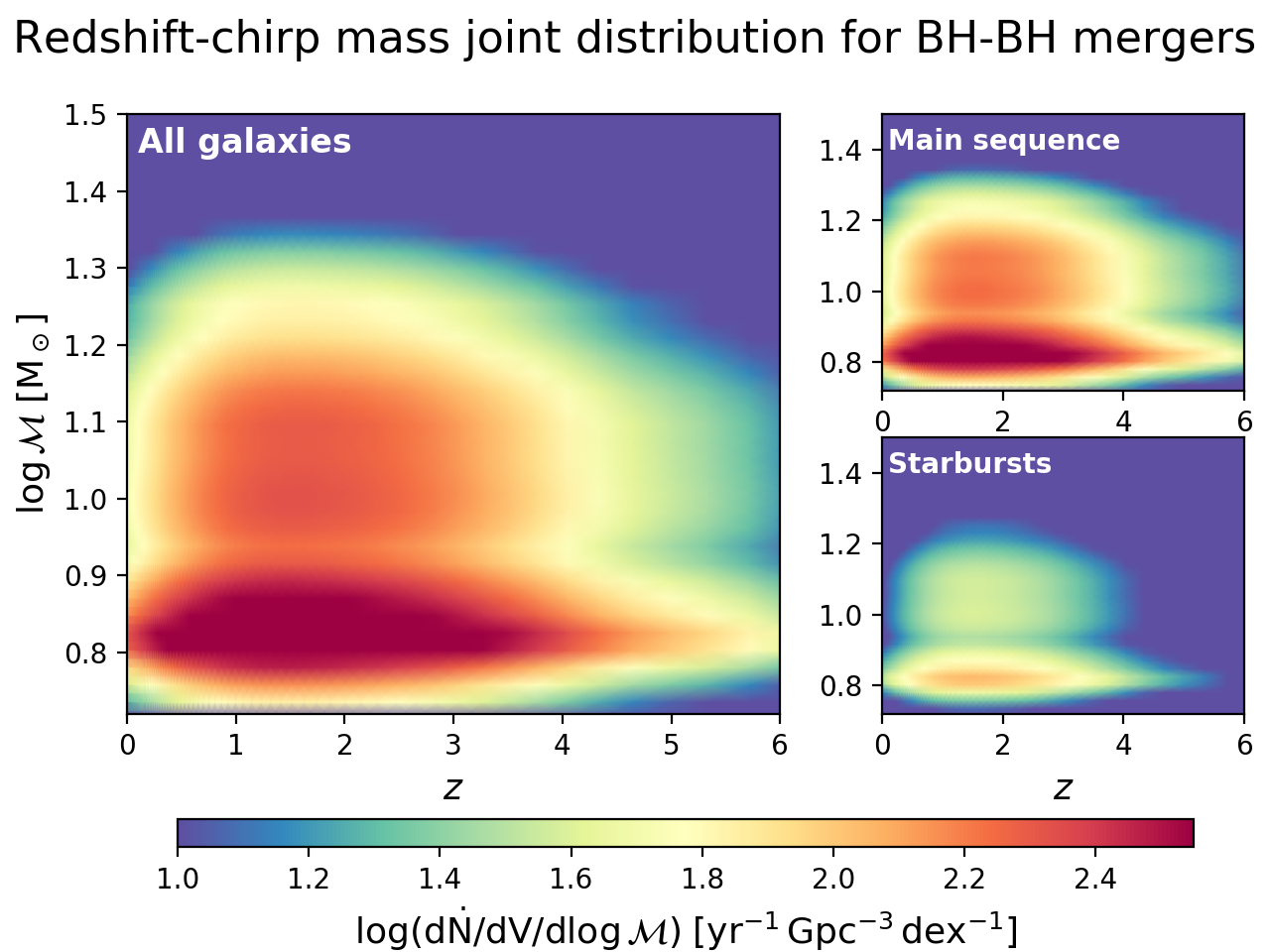}\\
    \caption{Top panel: merger rate density of double compact objects binaries as a function of redshift, computed using the GSMF as galaxy statistics and the FMR, following Eq.\eqref{eq:sfrd_FMR}. Blue lines refers to BH-BH, red lines to NS-NS, green lines to BH-NS events. Solid lines represents the merging happening in all the galaxies (main sequence and starbursts), while the dashed lines highlight the contribution of starbursts. The red and blue patches and the green arrow at $z\sim 0$ represents the LIGO/Virgo $90\%$ confidence intervals on the local rates for NS-NS and BH-BH and the upper limit for BH-NS after the O1, O2 and first half of O3 runs (Abbott et al. 2019; Abbott et al. 2020). Bottom panels: differential merging rates $\log{(\rm d\dot{N}/dV\,d\log\mathcal{M})}$ for the BH-BH case (color code) as a function of redshift and chirp mass. Contribution coming from all the galaxies (left panel), from main sequence galaxies (top right panel) and starbursts (bottom right panel).}
    \label{fig:contour_zm_FMR}
\end{figure*}

\begin{figure*}
    \centering
    \includegraphics[width=0.5\textwidth]{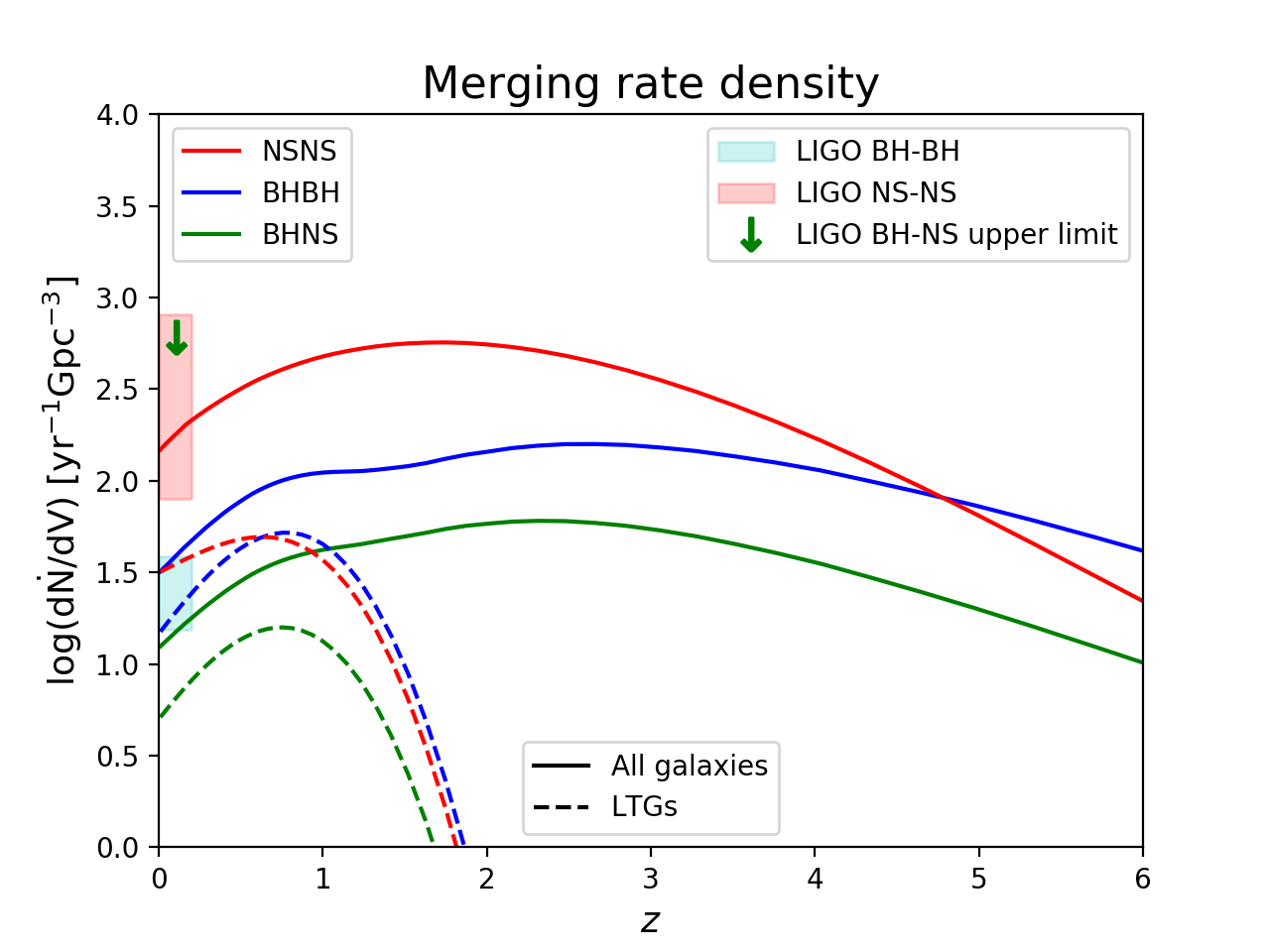}
    \includegraphics[width=0.5\textwidth]{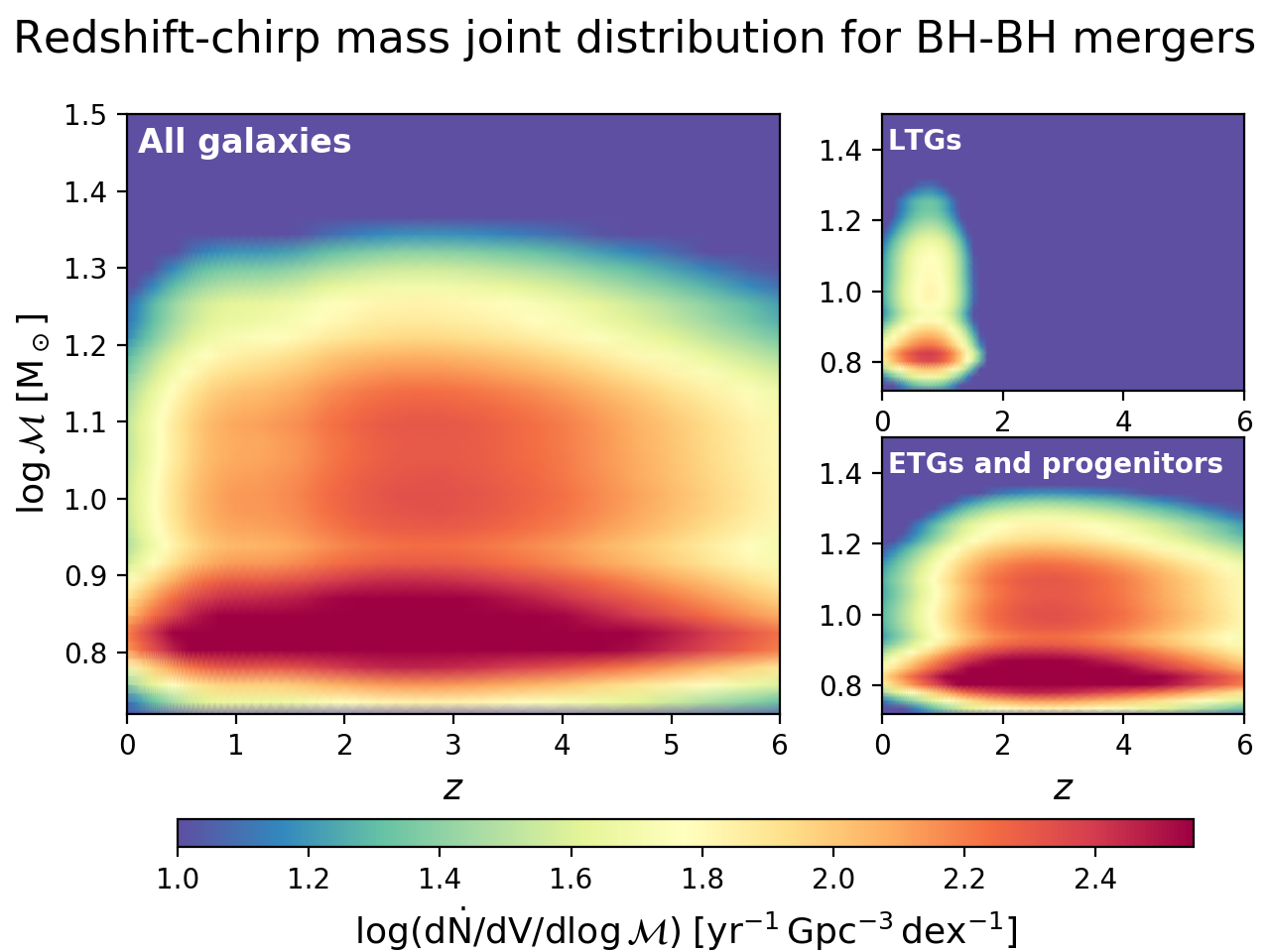}\\
    \caption{Top panel: merger rate density of double compact objects binaries as a function of redshift, computed using the SFRF as galaxy statistics and the FMR, following Eq.\eqref{eq:sfrd_FMR_SFR}. Blue lines refers to BH-BH, red lines to NS-NS, green lines to BH-NS events. Solid lines represents the merging happening in all the galaxies (LTGs, ETGs and their progenitors), while the dashed lines highlight the contribution of LTGs. The red and blue patches and the green arrow at $z\sim 0$ represents the LIGO/Virgo $90\%$ confidence intervals on the local rates for NS-NS and BH-BH and the upper limit for BH-NS after the O1, O2 and first half of O3 runs (Abbott et al. 2019; Abbott et al. 2020). Bottom panels: differential merging rates $\log{(\rm d\dot{N}/dV\,d\log\mathcal{M})}$ for the BH-BH case (color code) as a function of redshift and chirp mass. Contribution coming from all the galaxies (left panel), from LTGs (top right panel) and ETGs and their progenitors (bottom right panel).}
    \label{fig:contour_zm_SFR+FMR}
\end{figure*}

\begin{figure*}
    \centering
    \includegraphics[width=0.5\textwidth]{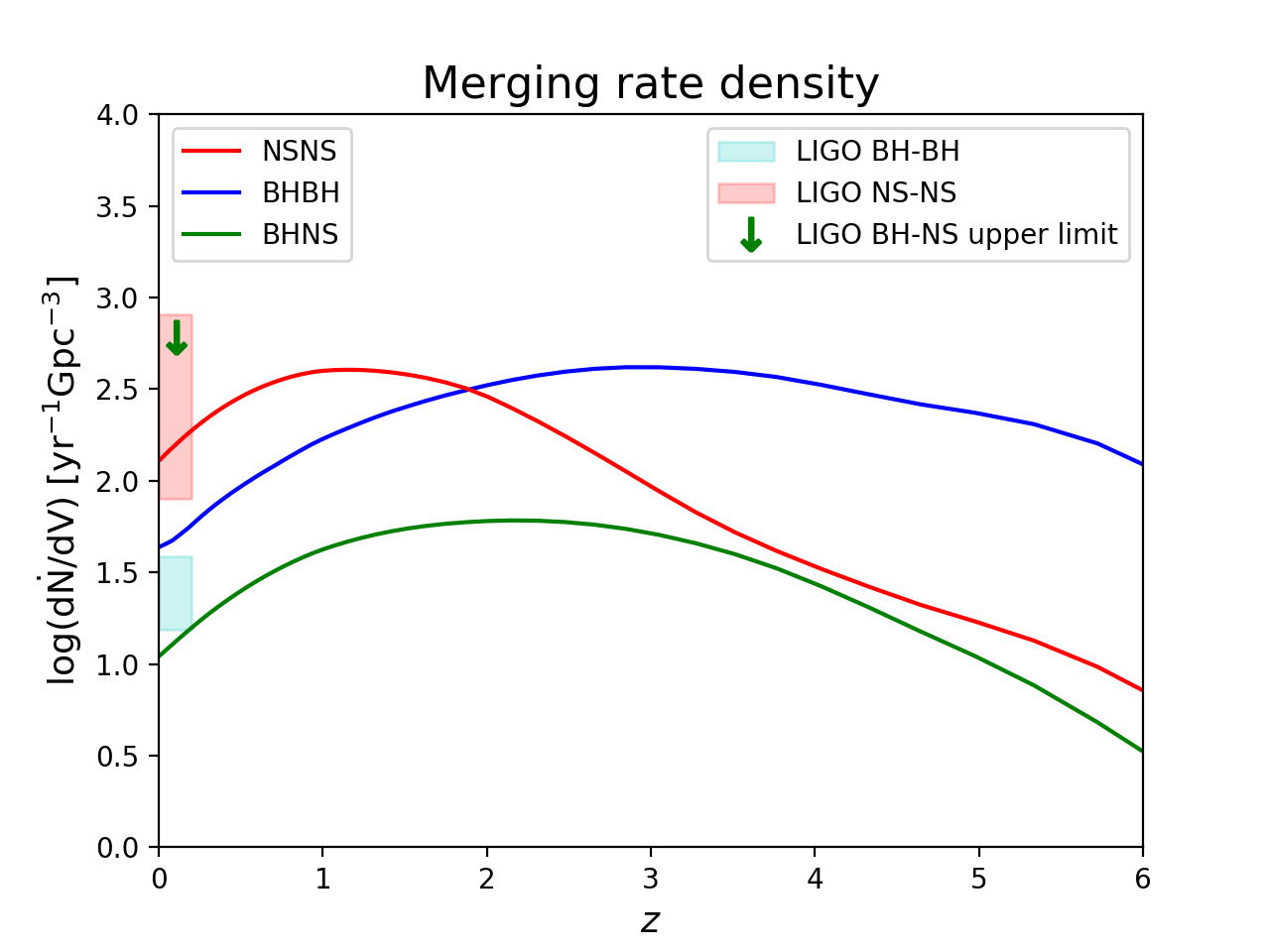}
    \includegraphics[width=0.5\textwidth]{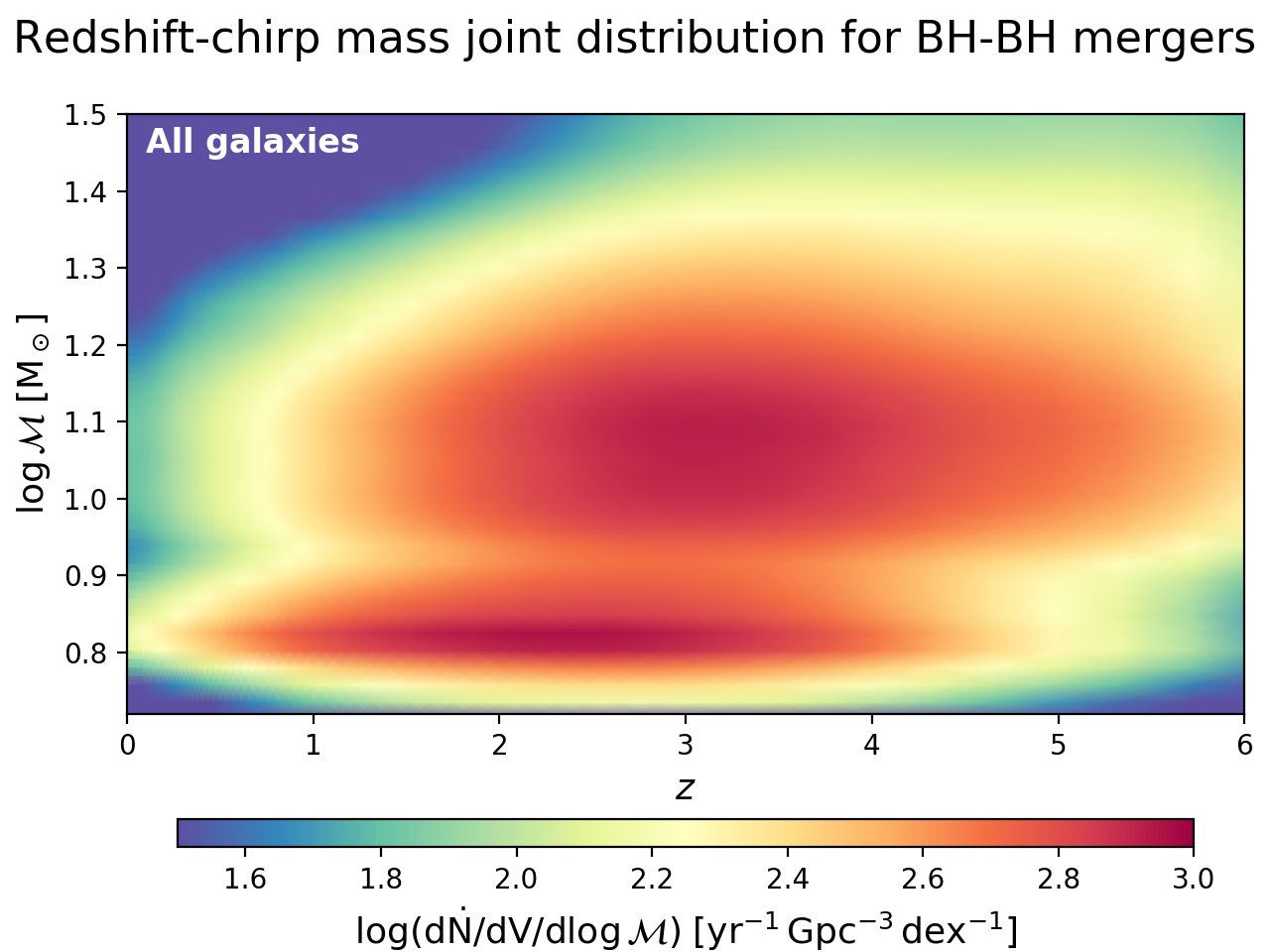}
    \caption{Top panel: merger rate density of double compact objects binaries as a function of redshift, computed using the GSMF as galaxy statistics and the MZR, following Eq.\eqref{eq:sfrd_MZR}. Blue lines refers to BH-BH, red lines to NS-NS, green lines to BH-NS events. The red and blue patches and the green arrow at $z\sim 0$ represents the LIGO/Virgo $90\%$ confidence intervals on the local rates for NS-NS and BH-BH and the upper limit for BH-NS after the O1, O2 and first half of O3 runs (Abbott et al. 2019; Abbott et al. 2020). Bottom panel: differential merging rate $\log{(\rm d\dot{N}/dV\,d\log\mathcal{M})}$ for the BH-BH case (color code) as a function of redshift and chirp mass. Note the change in the color code scale due to the larger number of BH-BH mergers occurring in this case.}
    \label{fig:contour_zm_MZR}
\end{figure*}

\begin{figure*}
    \centering
    \includegraphics[width=0.5\textwidth]{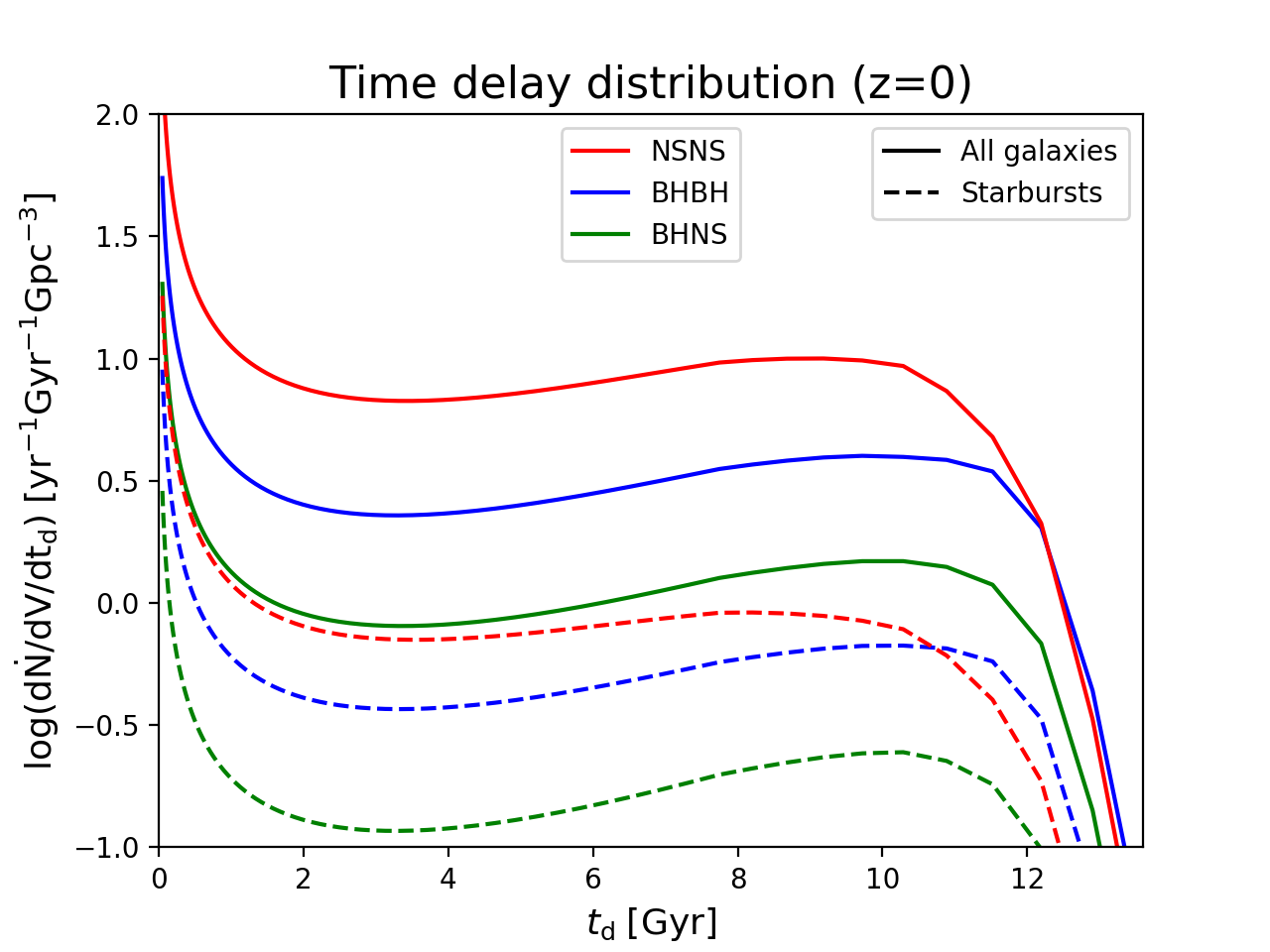}
    \includegraphics[width=0.5\textwidth]{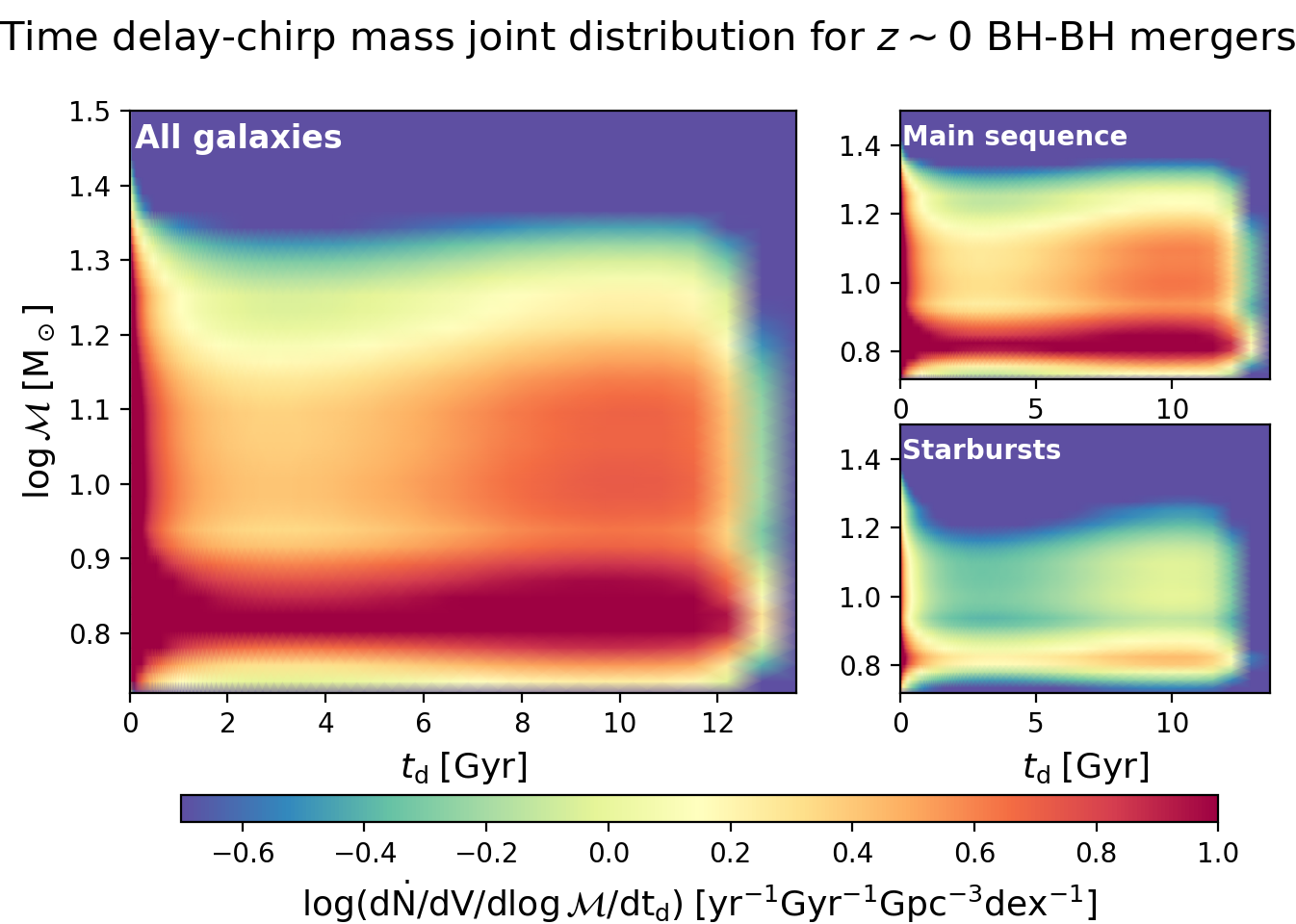}
    \caption{Top panel: differential merging rate $\log(\rm d\dot{N}/dV\,dt_d)$ at $z\sim 0$ for BH-BH as a function of the time delay between the formation of the binary and the merger, computed using the GSMF as galaxy statistics and the FMR, following Eq.\eqref{eq:sfrd_FMR}. Bottom panels: differential merging rates $\log(\rm d\dot{N}/dV\,d\log\mathcal{M}\,dt_d)$ at $z\sim 0$ for BH-BH as a function of the chirp mass and time delay. On the x axis there is the time delay, on the y axis the chirp mass and the color code represents the logarithmic number density of merging events. Contribution coming from all the galaxies (left panel), from main sequence galaxies (top right panel) and starbursts (bottom right panel).}
    \label{fig:contour_FMR_z0}
\end{figure*}

\begin{figure*}
    \centering
    \includegraphics[width=0.5\textwidth]{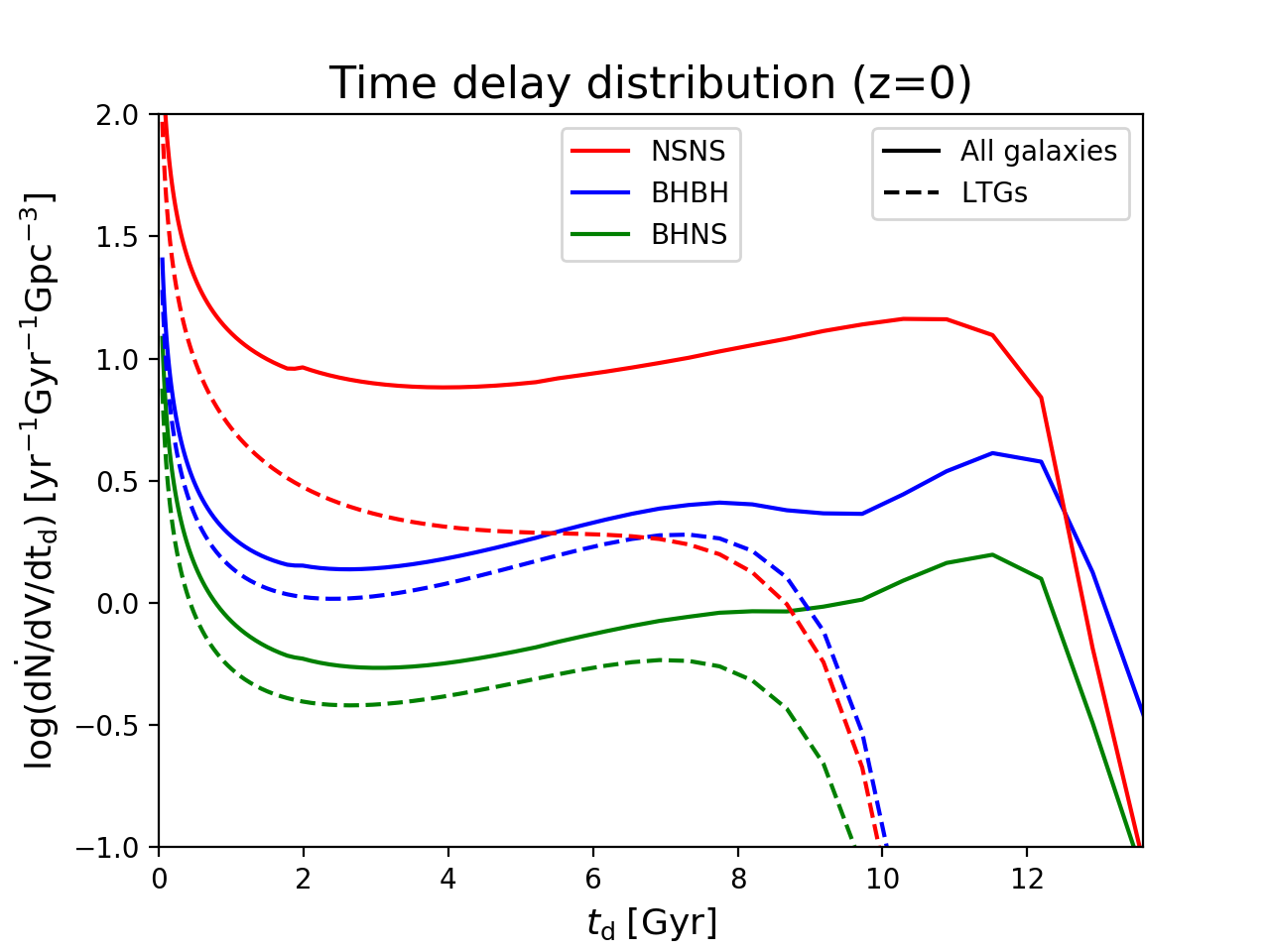}
    \includegraphics[width=0.5\textwidth]{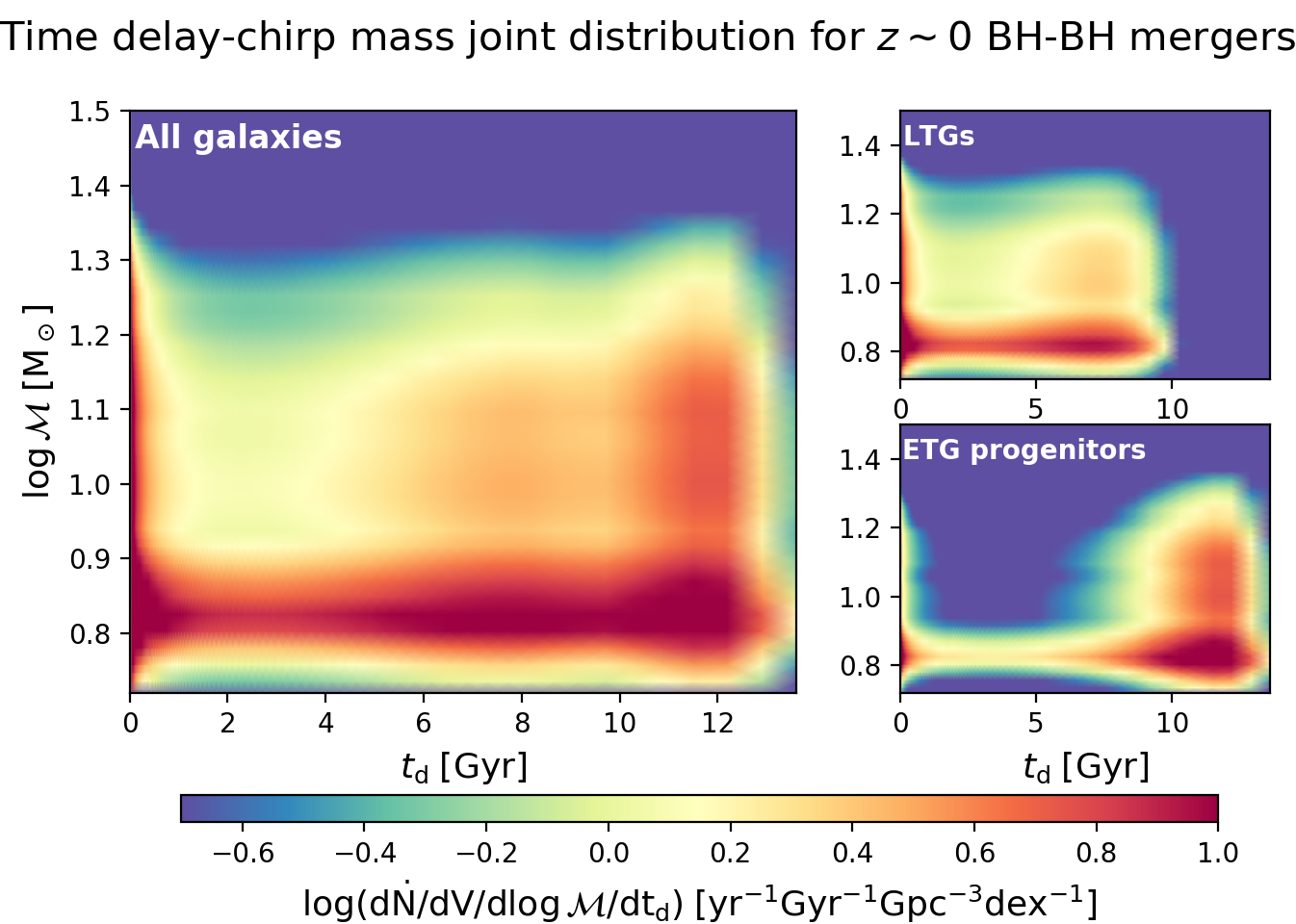}
    \caption{Top panel: differential merging rate $\log(\rm d\dot{N}/dV\,dt_d)$ at $z\sim 0$ for BH-BH as a function of the time delay between the formation of the binary and the merger, computed using the SFRF as galaxy statistics and the FMR, following Eq.\eqref{eq:sfrd_FMR_SFR}. Bottom panels: differential merging rate $\log(\rm d\dot{N}/dV\,d\log\mathcal{M}\,dt_d)$ at $z\sim 0$ for BH-BH as a function of the chirp mass and time delay. On the x axis there is the time delay, on the y axis the chirp mass and the color code represents the logarithmic number density of merging events. Contribution coming from all the galaxies (left panel), from LTGs (top right panel) and ETG progenitors (bottom right panel).}
    \label{fig:contour_SFR+FMR_z0}
\end{figure*}

\begin{figure*}
    \centering
    \includegraphics[width=0.5\textwidth]{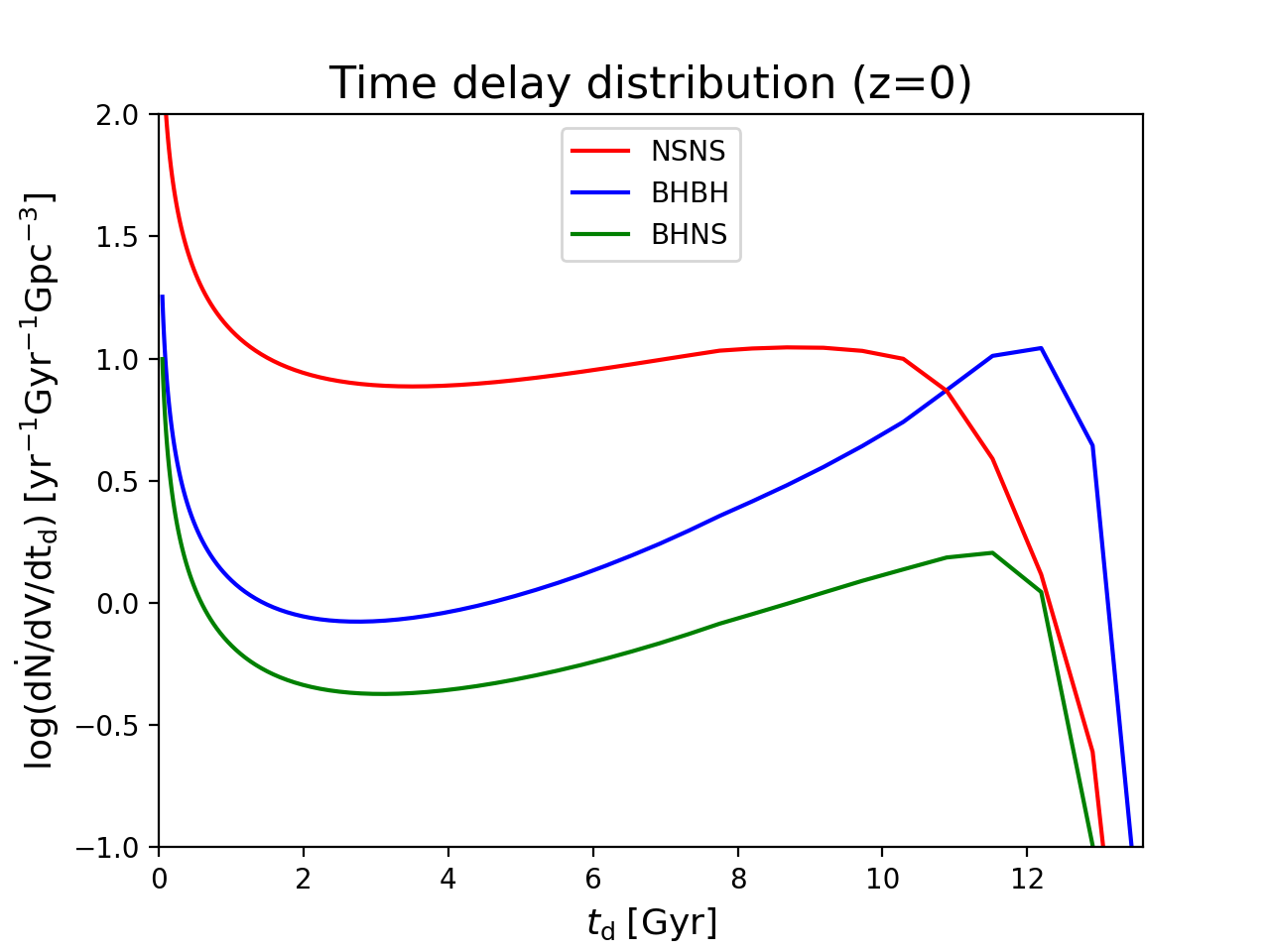}
    \includegraphics[width=0.5\textwidth]{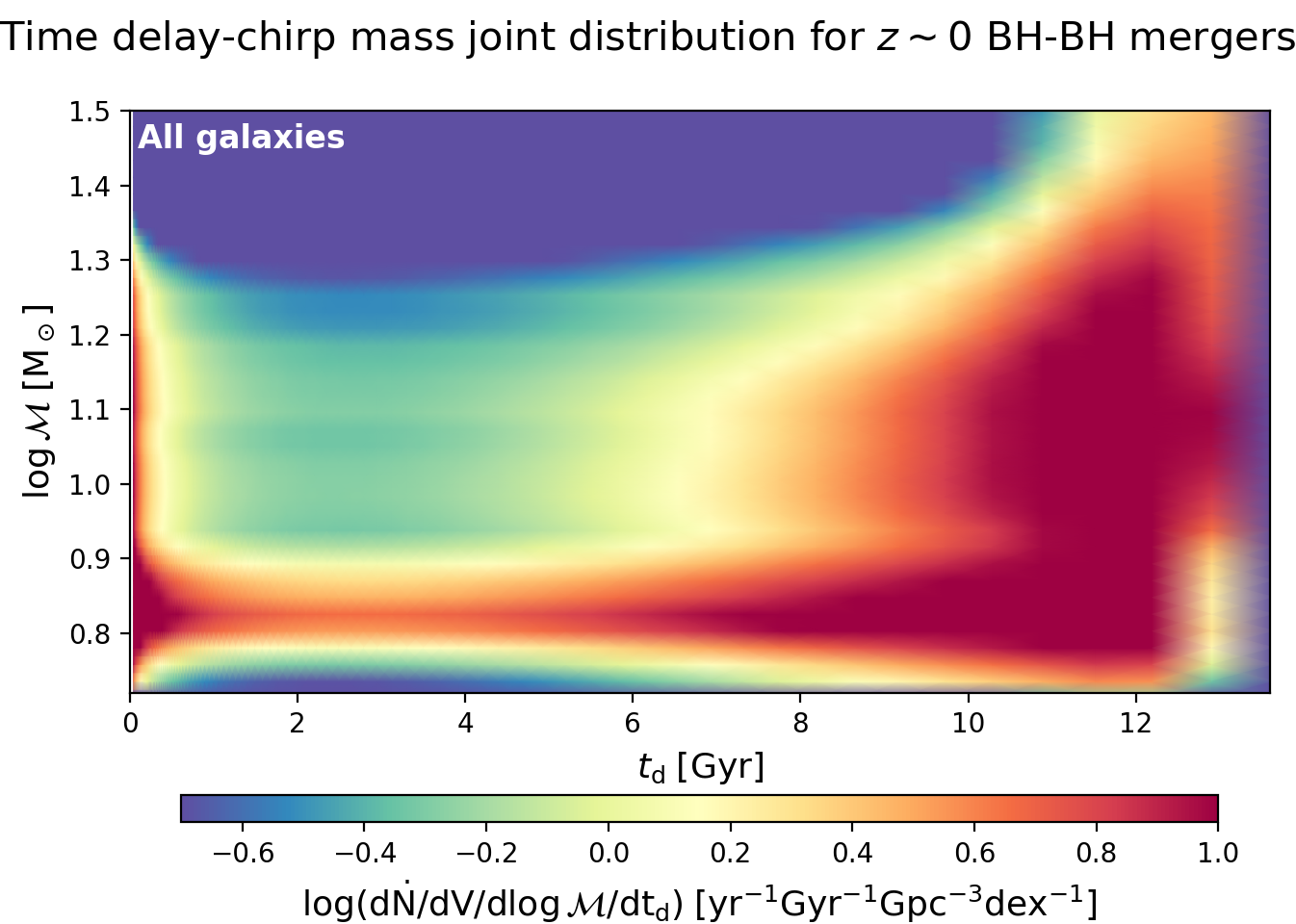}
    \caption{Top panel: differential merging rate $\log(\rm d\dot{N}/dV\,dt_d)$ at $z\sim 0$ for BH-BH as a function of the time delay between the formation of the binary and the merger, computed using the GSMF as galaxy statistics and the MZR, following Eq.\eqref{eq:sfrd_MZR}. Bottom panel: differential merging rate $\log(\rm d\dot{N}/dV\,d\log\mathcal{M}\,dt_d)$ at $z\sim 0$ for BH-BH as a function of the chirp mass and time delay. On the x axis there is the time delay, on the y axis the chirp mass and the color code represents the logarithmic number density of merging events.}
    \label{fig:contour_MZR_z0}
\end{figure*}

\subsection{The stellar term}\label{subsec:stellar term}
Within the isolated binary evolution scenario leading to the formation of merging DCOs (such as considered in this paper), the stellar term appearing in Eq. \eqref{eq:merging rates} is commonly obtained from binary population synthesis simulations (e.g. Belczynski et al. 2016, Eldridge et al. 2017, Mapelli et al. 2017, Stevenson et al. 2017). The outcome of those simulations (and therefore also the stellar term) depends on a number of assumptions made in order to describe the evolution of massive stars and binary interactions.
Many of those are highly uncertain (e.g. common envelope evolution, core-collapse physics and the related natal kicks) and are known to strongly affect the properties of the simulated populations of merging DCO (e.g. Portegies Zwart \& Yungelson 1998, Dominik et al. 2012, Chruslinska et al. 2018). 
Stellar evolution depends on metallicity - it affects, for instance, stellar winds and radii, also impacting the nature and outcome of binary interactions (e.g. Maeder 1992; Hurley et al. 2000; Vink et al. 2001; Belczynski et al. 2010a). As a consequence, the resulting stellar term also depends on metallicity.

This term can be separated into three main factors:
\begin{equation}
    \frac{\rm dN}{\rm dM_{\rm SFR}\,d\mathcal{M}\,dt_d}\,(Z)=\frac{\rm dN}{\rm dM_{\rm SFR}}\,(Z)\times\frac{\rm dp}{\rm d\mathcal{M}}\,(Z)\times\frac{\rm dp}{\rm dt_d}
\end{equation}
where $\rm dN/dM_{\rm SFR}$ is the number of merging DCOs formed per unit of mass formed in stars (formation efficiency)
at metallicity Z, $\rm dp/d\mathcal{M}$ is the metallicity-dependent chirp mass distribution
and $\rm dp/dt_d$ is the distribution of delay times between the formation of the progenitor binary and the DCO merger. 

The delay time distribution resulting from binary population synthesis is commonly found to be well described with a simple inverse proportionality $\rm dp/dt_d\propto t_d^{-1}$, independent of the DCO type or metallicity. We assume $\rm dp/dt_d\propto t_d^{-1}$  with the minimum $\rm t_{\rm d,min}$ = 50 Myr. The distribution is normalized to unity between $\rm t_{\rm d,min}$ and the age of the Universe.

We base the remaining two factors on the results of population synthesis calculations, using the model 'reference B' from Chruslinska et al. (2018) \footnote{We use the simulation data publicly available  under this url: \url{https://www.syntheticuniverse.org/}}. 
We note that this is just an example. The exact results concerning the populations of merging DCO presented further in this section would generally be affected by the choice of the population synthesis model (e.g. Chruslinska et al. 2019).
However, the main focus of this work is on the galactic term and the more in-depth discussion of the uncertainties related to binary evolution is beyond the scope of this study.

The formation efficiency as a function of metallicity for the chosen evolutionary model is shown in Chruslinska et al. 2019 (Fig. 1, thin lines).
The formation efficiency of merging BH-BH is typically found to show a strong low metallicity preference (e.g. Belczynski et al. 2010b; Dominik et al. 2012; Eldridge \& Stanway 2016; Stevenson et al. 2017; Klencki et al. 2018; Giacobbo et al. 2018). This dependence is also present in the chosen model, with BH-BH formation efficiency dropping by almost two orders of magnitude between 0.2 Z$_{\odot}$ and 0.4 Z$_{\odot}$. This dependence is generally weaker for other DCO types and in the adopted model shows a factor of $\lesssim$10 increase/decrease towards high metallicity for NS-NS/BH-NS.

Metallicity dependence of the chirp mass distribution is to large extent a consequence of the metallicity dependence of the maximum mass of the stellar remnant resulting from single stellar evolution. 
Due to metallicity dependent line-driven wind mass loss rates of massive stars (e.g. Vink et al. 2001, Vink \& de Koter 2005, Sundqvist et al. 2019, Sander, Vink \& Hartman 2020),
star with the same initial mass leaves a more massive stellar remnant at lower metallicity.
As a result, a population of DCO containing a higher fraction of objects originating from low metallicity progenitors will result in a chirp mass distribution with a more extended high mass tail (e.g. Fig. 4 in Chruslinska et al. 2019).

\subsection{Merging rates computation}\label{subsec:merging rates}

The merging rates per chirp mass units are computed as in Eq.\eqref{eq:merging rates} and, performing a further integration over the chirp mass, we get the redshift distribution of the merging rate density for the three types of merging binaries: BH-BH, NS-NS and BH-NS. In Figs. \ref{fig:contour_zm_FMR}, \ref{fig:contour_zm_SFR+FMR} and \ref{fig:contour_zm_MZR} we show, respectively, the results for the three different ways to compute the galactic term described in this paper: GSMF+FMR (Eq.\eqref{eq:sfrd_FMR}), SFRF+FMR (Eq.\eqref{eq:sfrd_FMR_SFR}) and GSMF+MZR (Eq.\eqref{eq:sfrd_MZR}). 

In the top panels are plotted the merging rates redshift distributions, highlighting the contribution of starburst galaxies in the GSMF+FMR case and the contribution of LTGs in the SFRF+FMR case. The local ($z\sim 0$) merging rates determinations by LIGO/Virgo for BH-BH ($15.3-38.8\,\rm Gpc^{-3}\,yr^{-1}$), NS-NS ($80-810\,\rm Gpc^{-3}\,yr^{-1}$) and BH-NS ($\leq 610\,\rm Gpc^{-3}\,yr^{-1}$) are also reported (see Abbott et al. 2019; Abbot et al. 2020b). The NS-NS and BH-NS merging rates fall inside the LIGO/Virgo interval for all the 3 cases considered, the BH-BH merging rates are slightly above the LIGO/Virgo interval for the GSMF+FMR and GSMF+MZR cases, while they fall inside for the SFRF+FMR case\footnote{We stress again that the agreement/disagreement with the LIGO/Virgo determinations can be due to the modelization of the stellar term. Moreover this is only one of the many constraint that a galactic or stellar model should be able to satisfy. So, the local rate alone does not represent a proof of the goodness of a model with respect to the others.}.  Comparing the top panels of Fig. \ref{fig:contour_zm_FMR} and \ref{fig:contour_zm_MZR}, referring to the GSMF+FMR and GSMF+MZR cases, which uses the same galaxy statistics as a starting point (GSMF) and differ only in the metallicity prescriptions, we can notice that the merging rates of NS-NS and BH-NS are similar, since they are less dependent on metallicity. On the other hand, the BH-BH merging rates, which are strongly dependent on metallicity, are substantially different: at low redshift ($z<1.5$) they are similar due to the similar behaviour of the metallicity distribution, while at high redshift ($z\geq 1.5$) they are larger for the MZR case (by a maximum factor of $\sim 10$) due to the strong decrease of metallicity at high redshift in the MZR case (see Figs. \ref{fig:GMF_FMRmannucci} and \ref{fig:GMF_MZR}). As for the SFRF+FMR case (Fig. \ref{fig:contour_zm_SFR+FMR}) the BH-BH merging rates lies in between. In fact, even if the metallicity, assigned through the FMR, stays rather high suppressing BH-BH mergers, this fact is partially compensated by the higher cosmic SFR density at high redshift obtained when the SFRF are employed as galaxy statistics (see Fig. \ref{fig:cosmic_sfrd}). This is also reflected in the larger merging rates for NS-NS and BH-NS in the SFRF+FMR case.

The bottom panels of Figs. \ref{fig:contour_zm_FMR}, \ref{fig:contour_zm_SFR+FMR} and \ref{fig:contour_zm_MZR} are also rather informative. They show the redshift and chirp mass distribution of the BH-BH merging rates. In the GSMF+MZR case (Fig. \ref{fig:contour_zm_MZR}) the chirp mass distribution extends up to $\mathcal{M}\gtrsim 30\,\rm M_\odot$ at high redshift ($z\geq 2$) where the metallicity tends to drop at subsolar values. This high chirp mass tail is reduced for the cases in which the FMR is used (Fig. \ref{fig:contour_zm_FMR} and \ref{fig:contour_zm_SFR+FMR}), since metallicity never drops too much, even at high redshifts, producing remnants with lower masses on average. We note here that none of the three cases analyzed is able to reproduce the high chirp mass events ($\mathcal{M}\gtrsim30\,\rm M_\odot$) recently observed at $z<1$ by the LIGO/Virgo collaboration (see Abbott et al. 2020a; Abbott et al. 2020b). This is not an issue for us, since even the chirp mass distribution, as well as the total rates of DCOs mergers, are strongly dependent on the selected model of stellar and binary evolution, whose discussion is out of the scope of the current work. The comparisons shown here are useful just to understand the general trend of DCOs mergers for different galactic prescriptions, but are not meant to reproduce the real chirp mass distribution. However, we stress that, in the three cases shown here, events with $\mathcal{M}\gtrsim30\,\rm M_\odot$ are still produced, simply their rate is much less than the rate of $\mathcal{M}<30\,\rm M_\odot$ events. A GW detector as AdvLIGO/Virgo would tend to detect mainly high chirp mass events since they produce stronger GW signals, so that the chirp mass distribution of merging DCOs may substantially be altered by selection effects. In addition, other channels of GW emission should not be excluded: dynamical formation and merger of compact object binaries (see e.g. Boco et al. 2020) as well as primordial black holes mergers (see e.g. Scelfo et al. 2018) could somewhat contribute to the GW detections and change the detected chirp mass distribution.

In the small plots on the bottom right of Figs. \ref{fig:contour_zm_FMR} and \ref{fig:contour_zm_SFR+FMR} it is shown the contribution of main sequence galaxies and starbursts, for the GSMF+FMR case (Fig. \ref{fig:contour_zm_FMR}), and the contribution of LTGs, ETGs and their star forming progenitors, for the SFRF+FMR case (Fig. \ref{fig:contour_zm_SFR+FMR}). Between main sequence galaxies and starbursts, no evident difference can be found, it is clear just that main sequence galaxies are the main contributors to the BH-BH merging rates. This is clearly dependent on the way we choose to model starbursts: we fixed their fraction to be $\sim3\%$ for all the stellar masses at all redshifts; it would be interesting to see how this would change treating the starburst fraction in a more detailed way (see Chruslinska et al. in prep.). Instead, the contribution of LTGs and ETGs is substantially different: LTGs contribute to the merging rates only at low redshift ($z\leq 2$) while only ETG progenitors are present at higher redshifts. At $z\leq 2$ the relatively longer tail towards larger chirp masses of LTGs can be explained by the fact that they have, on average, lower metallicities (see Fig. \ref{fig:SFR_FMR}).

\subsection{Chirp mass and time delay}\label{subsec:time delay}

The differential merging rates as a function of chirp mass and time delay tell us how the time delays and chirp masses are distributed for the merging events. They can be very helpful even for host galaxy association, since, given the chirp mass of the signal, they give informations on the average age of the stellar population producing the merger. 

They can be computed by just avoiding the first integration in Eq.\eqref{eq:merging rates}:
\begin{equation}
    \frac{\rm d\dot{N}}{\rm dV\,d\mathcal{M}\,dt_d}(\rm t,t_d)=\int\rm dZ\frac{\rm dN}{\rm dM_{\rm SFR}\,d\mathcal{M}\,dt_d}\,(Z)\frac{\rm d\dot{M}_{\rm SFR}}{\rm dV\,dZ}(\rm t-t_d)
    \label{eq:timedelay_chirpmass}
\end{equation}

We presents results for BH-BH mergers at $z\sim 0$. The results are shown in Figs. \ref{fig:contour_FMR_z0}, \ref{fig:contour_SFR+FMR_z0} and \ref{fig:contour_MZR_z0} for the three different ways to compute the galactic term described in this paper: GSMF+FMR (Eq.\eqref{eq:sfrd_FMR}), SFRF+FMR (Eq.\eqref{eq:sfrd_FMR_SFR}) and GSMF+MZR (Eq.\eqref{eq:sfrd_MZR}). 

The top panels illustrate the merging rates per units of time delay $\rm d\dot{N}/dV/dt_d$, meaning that Eq. \eqref{eq:timedelay_chirpmass} has been integrated over the chirp mass. In the GSMF+FMR case we show the contribution of starburst galaxies, while in the SFRF+FMR case we show the contribution of LTGs. In all the Figs. it is clearly visible a double peak distribution: the peak at low delay times is due to the shape of the intrinsic time delay distribution $\rm dp/dt_d\propto t_d^{-1}$ favouring short time delays, while the peak at $\rm t_d\sim 10-12\,\rm Gyr$ is due to the huge amount of star formation happening at redshift $z\sim 2-3$ that compensates for the time delay distribution favouring short time delays: a small fraction of the many objects formed at $z\sim 2-3$ can be seen through GW emission at $z\sim 0$. 

Apart from this shape shared by all the three cases, there are some differences among them that we explain in the following. In the cases where the GSMF is used as statistics (Figs. \ref{fig:contour_FMR_z0} and \ref{fig:contour_MZR_z0}) the NS-NS time delay distribution is similar since NS-NS mergers are almost independent on metallicity. The BH-BH time delay distribution is instead rather different: while for the GSMF+FMR case (Fig. \ref{fig:contour_FMR_z0}) the time delay distribution is flatter, with $\sim 48\%$ of the BH-BH merging with $\rm t_{\rm d}\leq 6\,\rm Gyr$, for the GSMF+MZR case the second peak is more pronounced, with only $\sim 20\%$ of BH-BH merging with $\rm t_{\rm d}\leq 6\,\rm Gyr$ and many events with $\rm t_{\rm d}\geq 9-10\,\rm Gyr$. This is due to the fact that, as already seen, in the MZR case the metallicity is much lower at high redshift, increasing the contribution to the $z\sim 0$ merging events from BHs formed at high redshift. For the BH-NS mergers the same effect, even if milder, can be noted. The starbursts contribution, shown only in the GSMF+FMR case, is subdominant. Comparing the NS-NS and BH-BH mergers in starbursts it can be noted a slight difference with respect to the all galaxies case. In fact, while the NS-NS contribution is always larger than the BH-BH one in the all galaxies case, if we restrict to starbursts the two contribution are roughly comparable, with BH-BH events being even dominant with respect to NS-NS for $\rm t_{\rm d}>11\,\rm Gyr$; this is due to the average lower metallicities of starbursts that slightly enhances the occurrence of BH-BH mergers. 

In the SFRF+FMR (Fig. \ref{fig:contour_SFR+FMR_z0}) case NS-NS have a similar shape to the other cases for $\rm t_d\leq 9-10\,\rm Gyr$, while there is an enhancement at larger time delays, due to the higher cosmic SFR at $z\geq 2$. The contribution of LTGs to the NS-NS merging rates follows the relative abundance of LTGs with respect to ETGs with the cosmic time. For the BH-BH mergers the shape is in between the GSMF+FMR and GSMF+MZR case: it can be seen a decrease at $1\leq\rm t_d\leq 6\,\rm Gyr$ and a moderate enhancement at $\rm t_d\geq 10\,\rm Gyr$, with a resulting $\sim 37\%$ of the BH-BH mergers having $\rm t_{\rm d}<6\,\rm Gyr$. The behaviour at small time delays can be explained by the rather high metallicity at low redshift, and the enhancement at high time delays is due to the larger amount of cosmic SFR, even if the metallicity remains pretty high. The contribution at low time delays comes almost exclusively from LTGs, which are less metallic, as seen in Fig. \ref{fig:SFR_FMR}, while events with large time delays come from ETGs that formed stars at higher redshifts, producing the second peak at $\rm t_{\rm d}>10\,\rm Gyr$.

The bottom panels show also the dependence on the chirp mass. It can be seen that in the GSMF+MZR case high chirp mass events tend to have huge time delays, while the distribution for the GSMF+FMR case is smoother. This means that in the GSMF+MZR case the GW events at $z\sim 0$ with $\mathcal{M}\geq 20\,M_\odot$ can be clearly linked to long delay times ($\gtrsim 10\,\rm Gyr$) and so to an older stellar population, while in the GSMF+FMR case the association between chirp mass and time delay is much less clear. The SFRF+FMR case lies in between. In the bottom right small panels of Figs. \ref{fig:contour_FMR_z0} and \ref{fig:contour_SFR+FMR_z0} it is shown, respectively, the contribution of main sequence galaxies and starbursts and of LTGs and ETGs. Between main sequence and starburst galaxies differences are not so evident, due to our treatment of the starburst population, while between LTGs and ETGs the difference is huge: as already seen the ETGs clearly contribute mostly to events with large time delays ($\rm t_{\rm d}>9\,\rm Gyr$) and LTGs to the events with $\rm t_{\rm d}<9\,\rm Gyr$. High chirp mass events can come from both the populations.

\section{Conclusions}\label{sec:conclusions}
Throughout the paper we focused on the computation of the cosmic SFR density per unit metallicity ($\rm d\dot{\rm M}_{\rm SFR}/\rm dV\,d\log Z$) with different prescriptions for the galaxy statistics and for the metallicity scaling relations. In particular:
\begin{itemize}
    \item we have shown the similarities and differences of using the stellar mass functions and the SFR functions as galaxy statistics, finding a good agreement between the two methods up to $z\sim 2$ and a larger cosmic SFR density at $z>2$ if the SFR functions are employed (by a maximum factor of $\sim 2.5$) (see Section \ref{sec:galaxy_statistics} and Fig. \ref{fig:cosmic_sfrd}). We have also discussed the main advantages and drawbacks of the two approaches: on the one hand the SFR functions provide a more direct statistics of the SFR of galaxies, on the other hand the GSMF can be more useful in the estimation of the galaxies' metallicities (see discussion at the end of Section \ref{sec:galaxy_statistics} and Section \ref{sec:metallicity}).
    \item we have presented the two main empirical scaling relations to associate metallicities to galaxies: the Mass-Metallicity Relation and the Fundamental Metallicity Relation. We analyzed the similarities and differences between the two relations, showing that the extrapolation of the FMR yields rather large average metallicity values ($Z\sim 0.4-0.5\,Z_\odot$) even at $z>2$, while the MZR usually implies very low metallicities ($Z<0.1\,Z_\odot$) at $z>2$. We have brought theoretical arguments and recent observational evidences testifying that the metallicity of high redshift dusty obscured star forming galaxies is rather large, arguing that, in order to reproduce those metallicities we should rely on extrapolations of the FMR or of a slowly evolving MZR (see Section \ref{sec:metallicity}).
    \item we have combined our fiducial scaling relation (the FMR) with both the two aforementioned galaxy statistics to compute the cosmic SFR density per units of metallicity ($\rm d\dot{\rm M}_{\rm SFR}/\rm dV\,d\log Z$) in the two cases (see Section \ref{sec:metallicity} and Figs. \ref{fig:GMF_FMRmannucci} and \ref{fig:SFR_FMR}). We have also considered an alternative case in which a sharply evolving MZR is used and combined it with the GSMF (see Fig. \ref{fig:GMF_MZR})). We find that the differences in the employed galaxy statistics and metallicity evolution are clearly reflected in the factor $\rm d\dot{\rm M}_{\rm SFR}/\rm dV\,d\log Z$.
\end{itemize}

Finally, in the last section, we have chosen a stellar and binary evolution model as an example to show the effect of the different galactic terms on the merging rates and on the properties of the merging binaries. We find that:
\begin{itemize}
    \item the merging rates computed with the different galactic terms are roughly consistent with the local merging rates determined by the LIGO/Virgo team (see Section \ref{sec:merging_rates});
    \item differences in the merging rate shape are present especially at high redshift where the two galaxy statistics and the two metallicity scaling relations are more different (see Section \ref{sec:merging_rates} and Figs. \ref{fig:contour_zm_MZR}, \ref{fig:contour_zm_FMR} and \ref{fig:contour_zm_SFR+FMR}). In short, using the SFRF enhances the compact remnants production at $z>2$ with respect to the GSMF case. The metallicity relation used affects mainly the BH-BH merging rates and, in particular, an higher metallicity in the early universe (obtained through the extrapolation of the FMR) hampers the BH-BH merging events even by a factor $\sim 10$;
    \item differences are present also in the chirp mass and time delay distributions (see Section \ref{sec:merging_rates} and Figs. \ref{fig:contour_zm_FMR}, \ref{fig:contour_zm_SFR+FMR}, \ref{fig:contour_zm_MZR}, \ref{fig:contour_FMR_z0}, \ref{fig:contour_SFR+FMR_z0} and \ref{fig:contour_MZR_z0}). In short if there is little metallicity evolution with redshift, the association between the chirp mass of the GW event and the redshift at which the merging occurs is less clear than in the case in which there is a strong metallicity evolution.
\end{itemize}
However, we remark again that the results on the merging rates are also dependent on the selected stellar model, so they should be intended just as a case study to compare the effects of the different galactic terms.

We hope that this work on different prescriptions for the galaxy statistics and metallicity can help in understanding the main properties of the merging binaries that will be detected through GW, especially with the future third generation detectors as Einstein Telescope. Knowing the effect of different galactic properties on the features of the merging binaries can be extremely helpful in order to better understand the star formation and galaxy evolution across cosmic time, when a large statistics of GW events will become available. From the observational point of view, a huge boost in the characterization of galaxies star formation rate and metallicity at increasingly high redshifts will come with the advent of JWST, which is suitable for measuring line diagnostics from galaxies across a broad range of redshifts, eventually in synergy with (sub)mm instruments like ALMA.

We thank the anonymous referee for useful comments. We acknowledge financial support from the EU H2020-MSCAITN-2019 Project 860744 ‘BiD4BEST: Big Data applications for Black hole Evolution STudies’ and from the PRIN MIUR 2017 prot. 20173ML3WW 002, ‘Opening the ALMA window on the cosmic evolution of gas, stars and massive black holes’. L.B. warmly thanks Luigi Bassini, for many helpful discussions.
MC acknowledges support from the Netherlands Organisation for Scientific Research (NWO).

\end{document}